\documentclass{article}

\usepackage{arxiv}

\usepackage[utf8]{inputenc} 
\usepackage[T1]{fontenc}    
\usepackage{hyperref}       
\usepackage{url}            
\usepackage{booktabs}       
\usepackage{amsfonts}       
\usepackage{nicefrac}       
\usepackage{microtype}      
\usepackage{lipsum}		
\usepackage{graphicx}
\usepackage{natbib}
\usepackage{doi}
\usepackage{framed,multirow}

\usepackage{xcolor}
\definecolor{newcolor}{rgb}{.8,.349,.1}

\usepackage{amsmath}
\usepackage{diagbox}
\usepackage{lmodern}
\usepackage{siunitx}
\usepackage{booktabs}
\usepackage{etoolbox}
\usepackage{subfigure}
\usepackage{placeins}

\title{Physics informed neural network for charged particles surrounded by conductive boundaries}

\author{ \href{https://orcid.org/0000-0002-9140-3950}{\includegraphics[scale=0.06]{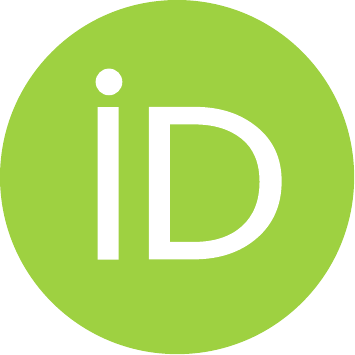}\hspace{1mm}Fatemeh Hafezianzade}\\
	Department of Physics\\
	Institute for Advanced Studies in Basic Sciences\\
	Zanjan, 45195-1159, Iran \\
	\texttt{fahafe98@gmail.com} \\
	\And	
	\href{https://orcid.org/0000-0001-5680-8066}{\includegraphics[scale=0.06]{orcid.pdf}\hspace{1mm}Morad Biagooi} \\
	Intelligent Data Aim Ltd (IDA Ltd)\\
	 Science and Technology Park of Institute for Advanced studies in Basic Sciences\\
	 Zanjan 45137-65697, Iran\\	 
	\And
	\href{https://orcid.org/0000-0002-7006-382X	}{\includegraphics[scale=0.06]{orcid.pdf}\hspace{1mm}SeyedEhsan Nedaaee Oskoee\thanks{Corresponding author}} \\
	Department of Physics\\
	Institute for Advanced Studies in Basic Sciences\\
	Zanjan, 45195-1159, Iran \\
	\texttt{nedaaee@iasbs.ac.ir} \\	
}

\renewcommand{\shorttitle}{\textit{arXiv} Template}

\begin{document}
\maketitle

\begin{abstract}
In this paper, we developed a new PINN-based model to predict the potential of point-charged particles surrounded by conductive walls. As a result of the proposed physics-informed neural network model, the mean square error and $R^{2}$ score are less than $7\%$ and more than $90\%$ for the corresponding example simulation, respectively. Results have been compared with typical neural networks and random forest as a standard machine learning algorithm. The $R^{2}$ score of the random forest model was $70\%$, and a standard neural network could not be trained well. Besides, computing time is significantly reduced compared to the finite element solver.
\end{abstract}


\section{Introduction}\label{sec1}

Computational Electromagnetic Simulation (CES) plays a significant role in many areas of science and engineering, such as soft matter, electrical engineering, biomedical engineering and chemistry. In addition, it has numerous applications in industry. For example, it is one of the main tools in investigating and designing the process of supercapacitors, which are porous energy storage devices with many applications in industry, especially when high power consumption or transfer is needed\cite{miller2008electrochemical}. Here, studying the physical mechanisms arising from charge storage in supercapacitors is essential for further technological development\cite{salanne2016efficient, Simon2008}.

Solving Maxwell's equation, especially the Poisson equation in this study, is an essential part of computational electromagnetic
algorithms\cite{Jackson1962}. Solving the Poisson equation can help scientists to calculate the potential of electrical sources in any system. 
However, many difficulties arise in practice due to the long-range nature of electrical interactions.  In particular, estimating the potential of point-charged components in an environment with conductive walls is challenging because of the induced charges presented on the boundaries.

Generally, there are two approaches to solving the Poisson equation: analytical solution\cite{Jackson1962} and numerical methods. There are limited techniques for solving analytically, like image charges methods applicable for cases with regular geometries;
however, there is no guarantee to achieve practical results. If, for example, a particle is placed in a cubic conductive container, the image charges method will produce an infinite series. On the other hand, numerical methods lead to approximate solutions based on discretizing space and/or time domains. One of the typical numerical methods is the Finite Element Method (FEM)\cite{Jin2014} which discretizes the continuous partial differential equations (PDEs) and forms a linear set of algebraic equations\cite{GeneH}. 
Nevertheless, even FEM fails in calculating the potential in a charged particle's position since the electrical potential is singular at the place of charges. There are a number of methods and algorithms that have been developed to address this problem, including Induced Charge MMM2D (ICMMM2D)\cite{Tyagi2007} for 2D, ELCIC\cite{Tyagi2008} for 2D + h, Induced Charge Computation ($ICC{*}$ )\cite{Tyagi2010,Kesselheim2010,Arnold2013} for 3D periodicity, and a method introduced by Reed et al.\cite{Reed2007} have been developed. In addition, recently, there has been another algorithm named PLT. It was first demonstrated for a partially periodic system constrained between two metallic plates in \cite{rostami2016highly}, and then it was applied to CAVIAR \cite{Biagooi2020}, a molecular dynamics simulation package for charged particles surrounded by non-trivial conductive boundaries.
Numerical solving of these problems with the CAVIAR package is accurate; moreover, it took less time than $ICC{*}$ \cite{Biagooi2020} but is still time and memory-consuming.

Recently another data-driven approach to solving the PDEs based on deep machine learning is also of great current interest. 
For instance, Shan et al. \cite{shan2020} present a CNN to predict the electric potential with different excitations and permittivity
distribution in 2D and 3D models. It is fast and efficient compared with FEM\cite{Jin2014}. However, a couple of problems prevent it from 
utilizing as a Poisson solver in the MD simulation process; first, it could not work in the case of discrete density functions such as those 
of point charges, and second, it is a physics-free approach which makes it hard to consider boundary conditions. 
To overcome the first problem, one can use the PLT algorithm. Additionally, Raissi et al. introduced the physics-informed neural network (PINN) that the loss function defined by \cite{Raissi2019} is an excellent alternative to the conventional deep learning method because of the governing equations, boundary conditions, and initial conditions used in its definition. 

In this paper, we applied a new PINN-based model to predict the potential of point-charged particles surrounded by conductive walls. We then compared the results with typical neural networks and random forests as a standard machine learning algorithm. For instance, we tried to implement these models for a charged particle in a spherical container. The reason for utilizing this simple example was that there is an exact solution to this problem through the analytical method, the image charges method. As a starting point, we used the PLT algorithm to transfer the Poisson equation into the Laplace equation with modified boundary conditions. Then we trained the model to solve the Laplacian equation with new boundary conditions. The input data is included the position in which we want to evaluate the potential on it and the modified boundary conditions; the output data is the corresponding electrical potential of that position.

\section{Methods}\label{sec2}
 
This article aims to build a machine-learning model (ML-Model) to predict the potential of point-charged particles surrounded by conductive walls. The potential of charged particles is calculated by solving the Poisson equation, which can be written as \cite{Jackson1962}:
\begin{equation}
    \nabla^{2} \phi=-\rho / \epsilon_{0}=-\sum_{i=1}^{N} q_{i} \delta(x-x_{q_{i}})/ \epsilon_{0},
    \label{eq1}
\end{equation}
where $\phi$ is the potential and $\rho$ is a charge distribution. The first and straightforward ML-Model that jumps to mind is a model that includes $x_{q}$ and $x$ as an input, and $\phi(x_{q},x)$ as an output. Here $x_{q}$ is the position of point-charged particle, $x$ is the position in which we want to calculate the potential on it, and $\phi(x_{q},x)$ is the corresponding potential. So the number of input features depends on the number of charged particles; for instance, in 3 dimensions, if there is N charged particles, the input features have to be $3 + 3\times N$. Therefore, this kind of model could only predict the potential of fix number of charged particles. In many applications of this method, such as the molecular dynamic simulation, this number is not fixed and could even increase or decrease during the simulation. We use the PLT algorithm to transpose the Poisson equation into the Laplace equation with new boundary conditions to overcome this problem. This algorithm will be discussed in more detail in \ref{subsec1}. So we can train a model which includes $x$ and modified boundary conditions as input features and $\phi(\phi_{b},x)$ as an output. We define the boundary conditions only on $N_{b}$ fixed points on the boundary $\left\{\phi_{1}, \phi_{2}, \ldots, \phi_{N_{b}}\right\}$. In this case, with the PLT algorithm, we can build a model with a fixed number of input features that can predict any charged particles' potential.
 
\begin{figure}[!htbp]
\centering
\includegraphics[trim={0cm 3cm 5cm 5cm},clip,scale=0.6]{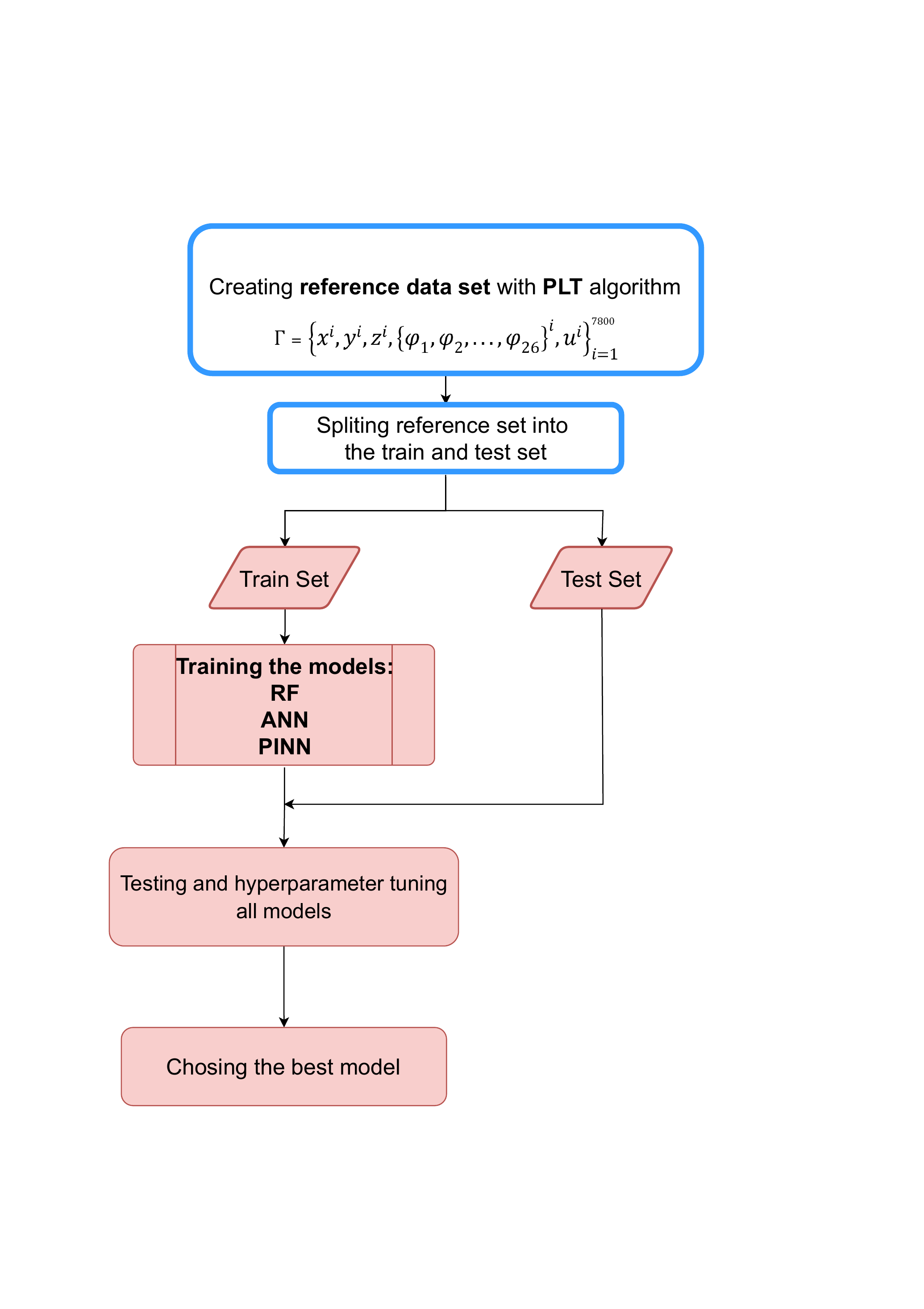}
\hrule
\caption{Methodology flow chart, The blue part: Preparing the data, in which the reference data set is created based on the PLT algorithm. The red part: Training models process, first the reference set is split to train and test set, then RF, ANN, and PINN model were applied on the train set, after tuning the hyperparameters the best model were chose. }\label{fig1}
\end{figure}
\FloatBarrier

\subsection{ Poisson to Laplace Transformation (PLT)}\label{subsec1}

According to the PLT algorithm, the electrical potential is divided into two parts: singular potential { $\left(\phi_{si }\right)$ } and smooth potential $\left(\phi_{sm}\right)$; {$\phi(\vec{x})=\phi_{s i}(\vec{x})+\phi_{s m}(\vec{x})$}. It is important to note that the smooth part here is the solution of the Laplace equation with modified boundary conditions,

\begin{equation}
    \begin{split}
        \nabla^{2} \phi_{s m}(\vec{x})=0,
        \label{eq2}
    \end{split}
\end{equation}
while { $\phi_{si }$ } obeys the famous Columb laws
\begin{equation}
\begin{split}
\phi_{s i}(\vec{x})=\sum_{i=1}^{N} \frac{q_{i}}{4 \pi \epsilon_{0}\|\vec{x}-\vec{x}_{i}\|}.
 \label{eq3}
 \end{split}
\end{equation}
It can be seen that the modified boundary condition for { $\phi_{sm }$ } is represented by 
\begin{equation}
\begin{split}
\phi_{s m}\|_{\vec{x}_{b c}}=\phi\|_{\vec{x}_{b c}}-\phi_{s i}\|_{\vec{x}_{b c}},
\label{eq4}
\end{split}
\end{equation}
where {$\phi\|_{\vec{x}_{b c}}$} corresponds to the initial electrical potential on the boundaries. Finally, with the PLT algorithm, we could transfer the Poisson to the Laplace equation with new modified boundary conditions, then train an ML-Model with these modified boundary conditions as an input parameter and the smooth potential as an output. Afterward with the summation of singular and predicted smooth potential, we can reach the total potential. Advantage of utilizing PLT algorithm is that it leads to having a fixed number of input data since the number of input data would be independent of the number of point-charged particles.

\begin{figure}[!htbp]
    \centering
    \subfigure[]{\includegraphics[width=0.3\textwidth,trim={3cm 3cm 0cm 0},clip,scale=0.75]{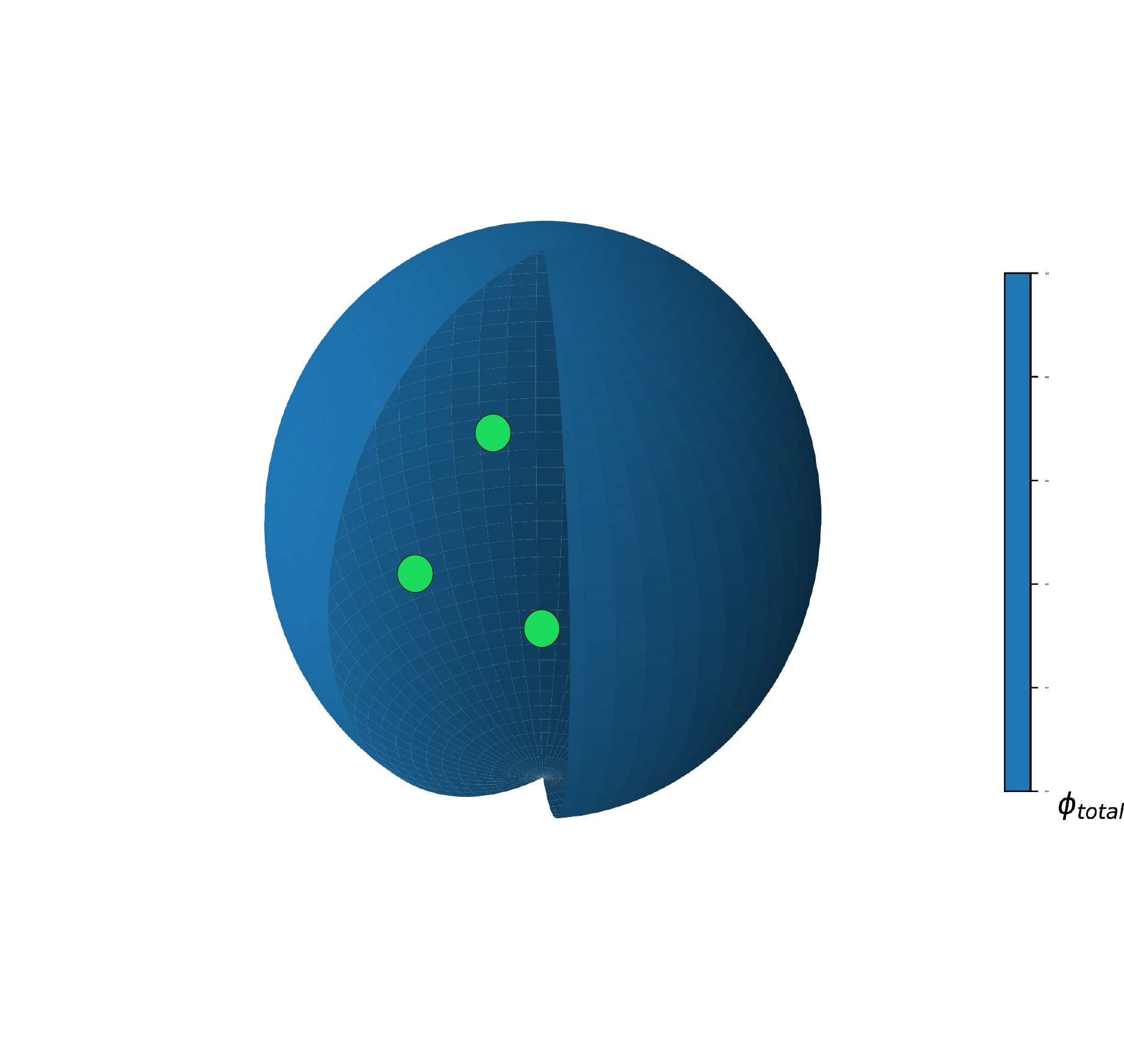}}
    \subfigure[]{\includegraphics[width=0.3\textwidth,trim={5cm 6cm 3cm 10},clip,scale=1]{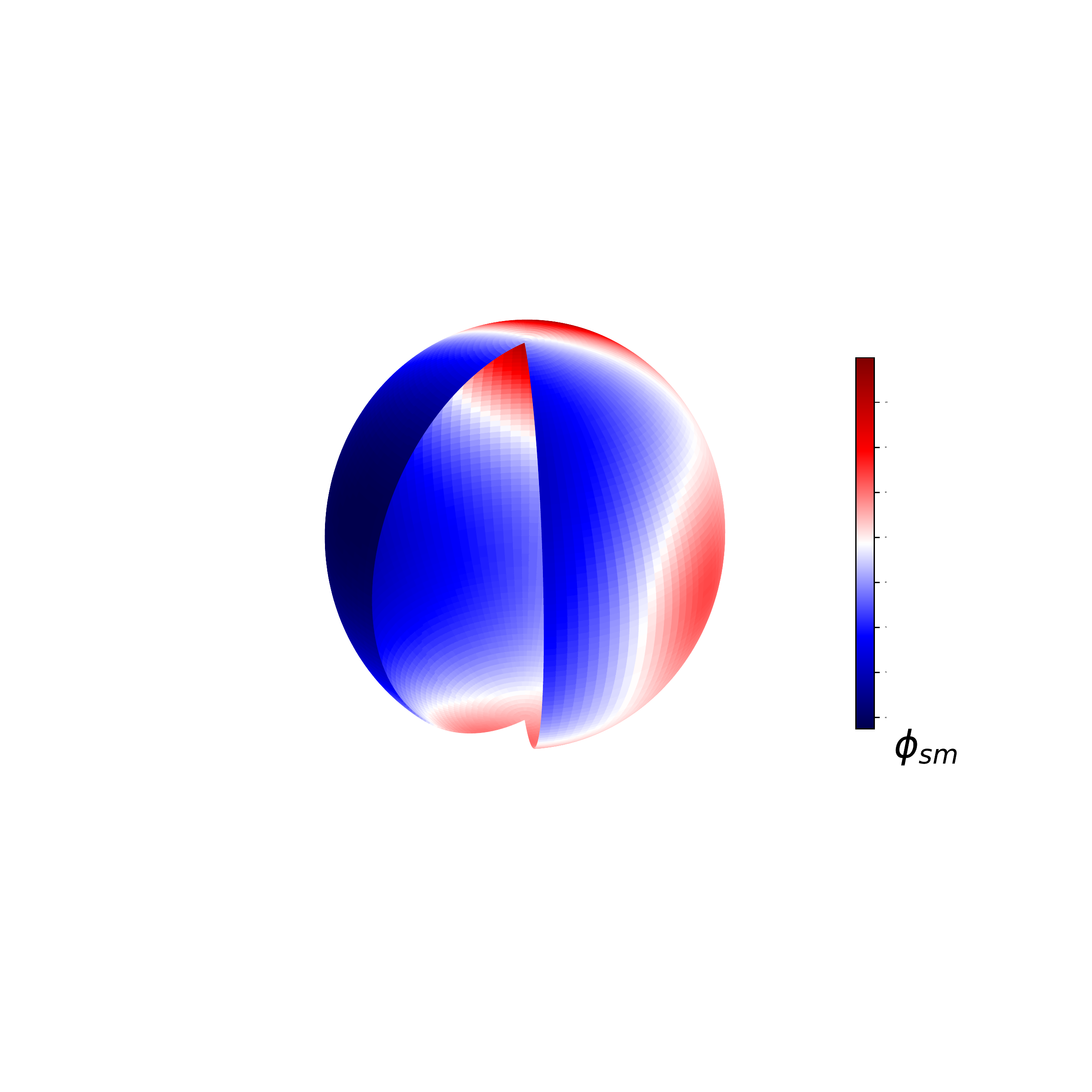}}
    \subfigure[]{\includegraphics[width=0.3\textwidth,trim={5cm 6cm 3cm 3},clip,scale=1]{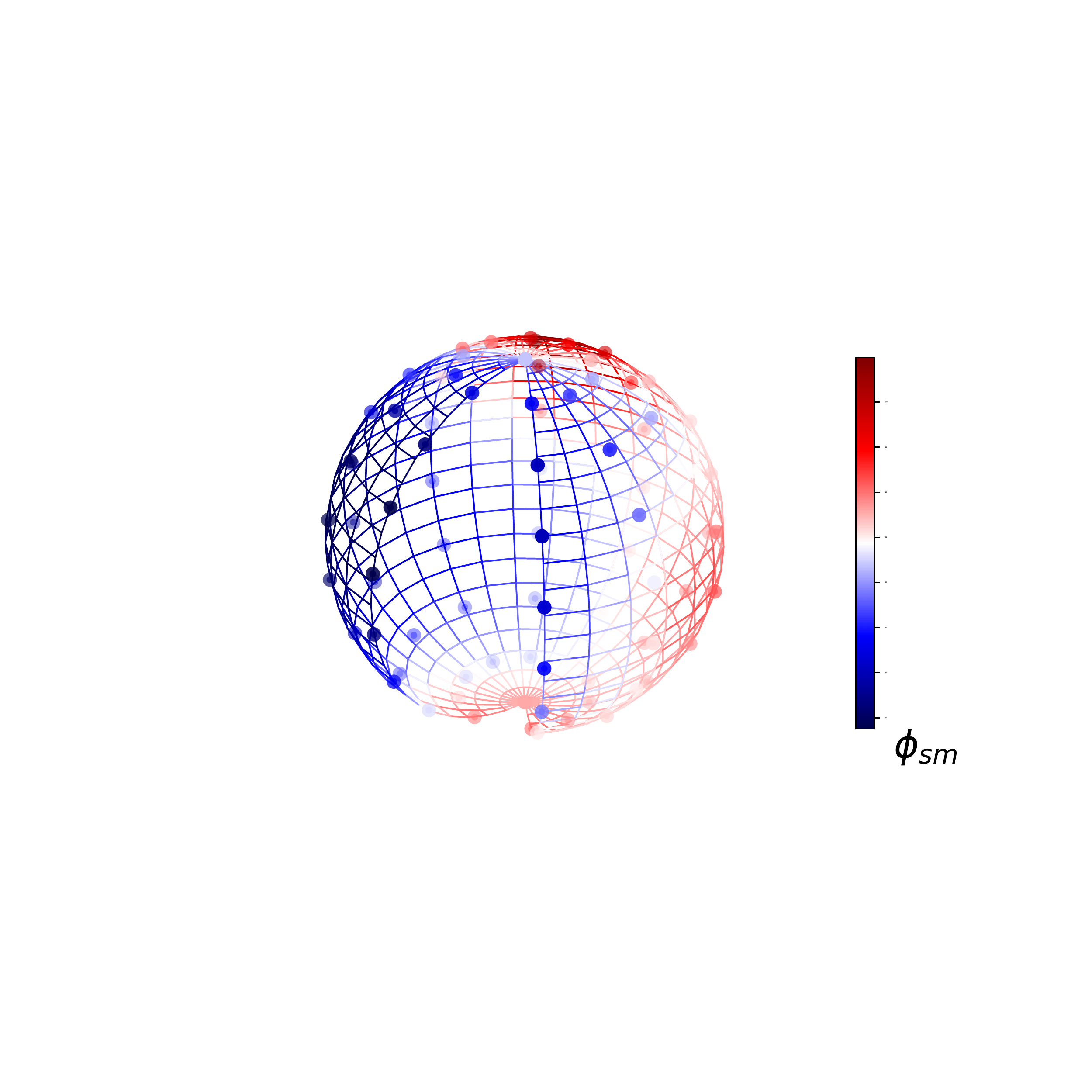}} 
    \caption{Schematic of the PLT method: a) the main system which had point charges inside of it b) the new system without any point charges and the boundaries were modified c) $N_{b}$ points on the boundary are shown to be used as our model input. }
    \label{fig4}
\end{figure}
\FloatBarrier

\subsection{Data Engineering}\label{subsec2}
For training a highly accurate model, having a nice train set is crucial. In this work, the reference set is $\Gamma=\left\{x^{i}, y^{i}, z^{i}, \vec{\varphi^{i}_{bc}} ,\phi^{i}\right\}_{i=1}^{N}$, where the input is concluded $\left\{x, y, z,\vec{\varphi}_{bc}=\left\{\phi_{1},\phi_{2},...,\phi_{N_{b}}\right\}\right\}$ and $\phi$ is the target. $x, y, z $ is the coordinate of a point in the container on which we want to calculate their potential, $\phi$ is the numeric value of the potential at this point, and $\left\{\phi_{1},\phi_{2},...,\phi_{N_{b}}\right\}$ is the boundary condition on $N_{b}$ points on the boundary. First, in the container, $N_{p}$ positions are chosen to predict the potential in their situations. In fact, for each boundary condition, $\left\{\phi_{1},\phi_{2},...,\phi_{N_b}\right\}$, there are $N_{p}$ points that we want to calculate the potential at their positions. Then, the reference set could be created for $N_{q}$ different boundary conditions. So the reference set consists of $N=N_{q}\times N_{p}$ samples which could split to train and test set.
In this case, our container is a sphere, we also set $N_{p}=78$, $N_{b}=26$, and $N_{q}=100$. So, our reference consists of 100 different boundary conditions and for each boundary condition $\left\{\phi_{1},\phi_{2},...,\phi_{26}\right\}$ there are 78 points in the sphere on which we want to calculate the potential on it. We use the solution of the image charges method (Eq.\ref{eq5}) to calculate targets of the reference set:

\begin{equation}
\begin{split}
\phi(\vec{x})=\frac{1}{4 \pi \epsilon_{0}}\{\frac{q}{\|\vec{x}-\vec{x}_{q}\|}+\frac{q^{\prime}}{\|\vec{x}-\vec{x_{q^{\prime}}}\|}\}, \quad q^{\prime}=-\frac{r q}{a}, \quad \vec{x_{q^{\prime}}}=\frac{a^{2}}{r} \frac{\vec{x}_{q}}{\|\vec{x}_{q}\|},
\label{eq5}
\end{split}
\end{equation}
where $a$ is the conductive spherical shell radius and $r$ is the distance of a point charge $q$ from its center. {The numeric value of potential is minimal ($\sim 10^{-9}$), which conducts to significant rounding error during computation; therefore, the potential of an electron in a $1m$ distance of it, $1.44 \times 10^{-9} [V]$, is used as a unit to make equation \ref{eq5} dimensionless}. We randomly chose 5000 and 1000 samples from the reference set to create a train and test set. The train and test set have no samples in common. In addition, the best model could adequately predict the potential of test samples and samples with different boundary conditions from the train and test set. So to evaluate the model better, we prepare an extrapolation set that includes 1000 samples with 55 distinct boundary conditions.

\subsection{ML Algorithms}\label{subsec3}
In this work, three different supervised learning methods have been used, and their regression accuracy, based on the metrics presented in \ref{subsec4}, has been evaluated. Mainly, we stick to Physics-Informed Neural Networks (PINN, \cite{Raissi2019}), but to compare our results with other ML algorithms, we use Random Forest (RF, \cite{Breiman2001}) and Artificial Neural Networks (ANN). All the models are briefly introduced, the hyperparameters are fine-tuned, and their performance is reported. Scikit-learn \cite{Pedregosa2011}, Tensorflow \cite{Abadi2016}, Keras \cite{Chollet2015}, and NumPy \cite{Van} are all the python libraries that have been used in this project.

\subsubsection{Random Forest(RF)}\label{subsub1}
RF is one of the most popular machine learning algorithms in regression problems for many reasons, but in this project, this model has been chosen since \\ a) It is speedy to learn. \\ b) It is robust against over-fitting.\\ over-fitting is detected when the performance of train samples is perfect while the performance of test samples is poor. RF is an ensemble model in which an average of many uncorrelated trees determines the predicted potential for the target data set. Although each tree is a weak learner, they make a strong learner when many trees are grouped. The RF randomizes the trees by choosing a subset of training data and features for each tree. Here we use scikit-learn \cite{Pedregosa2011} RF implementation.

\subsubsection{ANN}\label{subsub2}
Typical neural network architecture consists of the input layer, multiple hidden layers, and the output layer with several neurons in each layer. Totally:
\begin{itemize}
    \item Input layer: The neurons in the input layer are the input features.
    \vspace{0.25cm}
    \item Hidden layers: The value of every neuron in the hidden layers is a linear combination of the neurons in the previous layer followed by the implementation of an activation function(Eq.\ref{eq6}); in most cases, the activation function is non-linear.

    \begin{equation}
    {a}_{n}=\sigma_{l}\left(a_{n-1} \mathbf{w}_{n}+\mathbf{b}_{n}\right).
    \label{eq6}
    \end{equation}
    $\mathbf{n}$ is the layer number, $\mathbf{w}$ and $\mathbf{b}$ are the model parameters, weights  and bias respectively, and $\sigma_{l}$ is the activation function based on \cite{Goodfellow-et-al-2016}.
    \vspace{0.25cm}
    \item Output layer: The neurons in the output layer are the model targets and 
    they are calculated with Eq.\ref{eq6} with linear activation function.
    \vspace{0.25cm}
    \item Loss function: There is a function in all neural networks that must be minimized over the model parameters during the training stage via back-propagation, typically the loss function is the mean square error between the true and the predicted values.
    \vspace{0.25cm}
    
    \begin{equation}
    \operatorname{Loss}(w)= M SE_{d}+\lambda\sum_{w} w^{2},
    \label{eq7}
    \end{equation}   

    \begin{equation}
    M S E_{d}= \frac{1}{N}\sum_{\mathrm{i}=1}^{\mathrm{N}}\left[\mathrm{U}\left(\mathbf{X}_{\mathrm{i}}, \mathbf{w}\right)-\mathrm{T}_{\mathrm{i}}\right]^{2}.
    \label{eq7-2}
    \end{equation}   
\end{itemize}
U and T are the predicted output and true target values, respectively, X is the input data, and w is the parameter of neural networks, weights, and biases. The first sentence in Eq.\ref{eq7} is a mean square error, and The second sentence exists to prevent over-fitting, namely $L2$ regularization\cite{Connect1992}, that is used in order to reduce the effects of the large weights.
 
\subsubsection{PINN}\label{subsub3}
\begin{figure}
\centering
\includegraphics[trim={0 0cm 0 0cm},clip,width=\textwidth]{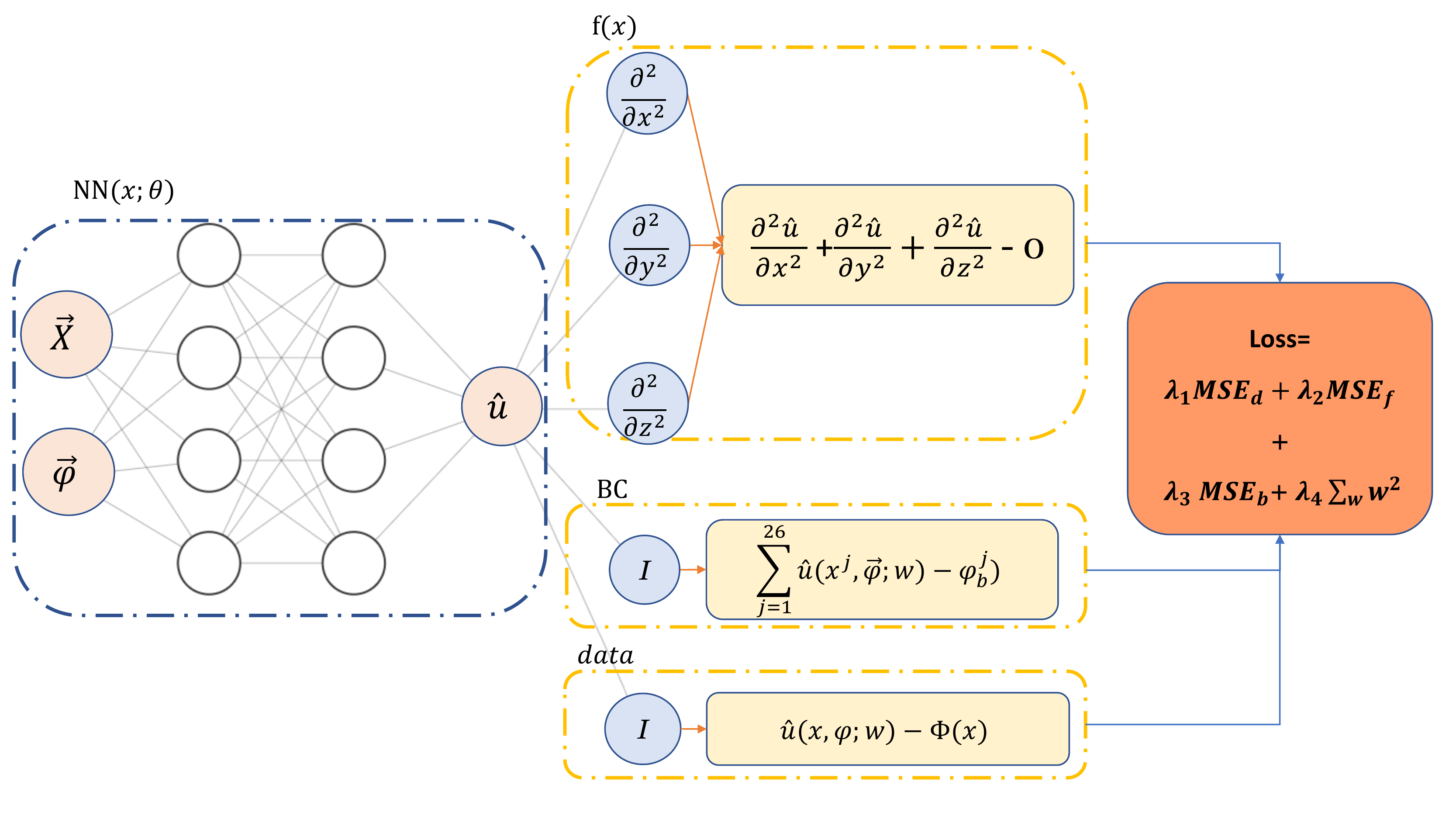}
\hrule
\caption{Physics-informed neural network scheme for solving Laplace equation with variable boundaries }\label{fig2}
\end{figure}

PINN\cite{Raissi2019} enforces the Laplace equation, a physical law of the electromagnetic system, as a constraint on the neural network. This study proposes a PINN-based approach to solve the Laplace equation with changeable boundary conditions. Fig.\ref{fig2} shows a schematic of the neural network layout for this approach. PINN-based models are neural networks with modified $loss$ functions: 

\begin{equation}
\operatorname{Loss}=\lambda_{1}M S E_{d}+\lambda_{2} M S E_{f} +\lambda_{3} M S E_{b} +\lambda_{4}\sum_{w} w^{2},
\label{eq10}
\end{equation}
The first and the last term of Eq.\ref{eq10} are the same as typical neural networks in Eq.\ref{eq7}. The second term corresponds to the governing physical equation, i.e., the is Laplace, and the third term corresponds to the boundary conditions;
 
\begin{equation}
M S E_{d}=\frac{1}{N_{d}} \sum_{i=1}^{N_{d}}\|u(\textbf{x}_{d}^{i},\vec{\varphi_{d}^{i}};w)-\phi_{d}^{i}\|^{2},
\label{eq11}
\end{equation}

 \begin{equation}
M S E_{f}=\frac{1}{N_{f}} \sum_{i=1}^{N_{f}}\|f(\textbf{x}_{f}^{i},u_{f}^{i};w)\|^{2},
\label{eq12}
\end{equation}
and
\begin{equation}
M S E_{b}=\frac{1}{N_{b}} \sum_{i=1}^{N_{b}}\|\mathcal{B} (\textbf{x}_{b}^{i},\vec{\varphi_{b}^{i}},u_{b}^{i};w)\|^{2}.
\label{eq13}
\end{equation}
\\
 Here we define $f\left(\textbf{x} ,\mathrm{u};w \right)$
\begin{equation}
\begin{split}
\begin{aligned}f\left(\textbf{x}, \mathrm{u};w\right)& =  0,\quad \quad \quad \mathrm{x}  \in \Gamma_{f} \\ & =  \nabla^{2} u \\ &=\frac{\partial^{2} u}{\partial x^{2}}+\frac{\partial^{2} u}{\partial y^{2}}+\frac{\partial^{2} u}{\partial z^{2}}
\\ &=\frac{\partial w}{\partial x}\frac{\partial}{\partial w}\left(\frac{\partial w}{\partial x}\frac{\partial u}{\partial w} \right)+\frac{\partial w}{\partial y}\frac{\partial}{\partial w}\left(\frac{\partial w}{\partial y}\frac{\partial u}{\partial w} \right)+ \frac{\partial w}{\partial z}\frac{\partial}{\partial w}\left(\frac{\partial w}{\partial z}\frac{\partial u}{\partial w} \right),\end{aligned}
\label{eq8}
\end{split}
\end{equation}
with Dirichlet boundary conditions
\begin{equation}
\begin{split}
\begin{aligned}\mathcal{B}\left (\textbf{x},\vec{\varphi},u;w\right)&=0 ,\quad\quad \quad\quad \quad \mathrm{x}  \in \Gamma_{b} \\ & = \sum_{j=1}^{26} (u(\textbf{x}^{j},\vec{\varphi};w) - \varphi_{b}^{j}).
\end{aligned}
\label{eq9}
\end{split}
\end{equation}
$\lambda_{1}, \lambda_{2}, \lambda_{3}$ in Eq.\ref{eq10} correspond to the weight coefficients for the data contributions, Laplace equation, and boundary losses. We use the weight coefficient by motivating from the study of Kag et al. \cite{kag2022physics}. The last sentence is the $L2$ regularization\cite{Connect1992}. Notice that the model with $\lambda_{2}=\lambda_{3}=0.0$ is exactly a typical neural network described in the previous subsection.

\subsection{Evaluating Metrics}\label{subsec4}
The performance evaluation of different algorithms for potential estimation depends on different metrics,
\begin{equation}
\begin{split}
\Delta \phi=\phi_{True}-\phi_{Pred},
\label{eq14}
\end{split}
\end{equation}

\begin{equation}
\begin{split}
\sigma=\sqrt{\frac{1}{n} \sum_{i=0}^{n-1}(\Delta \phi)^{2}},
\label{eq15}
\end{split}
\end{equation}

\begin{equation}
\begin{split}
\mathrm{R}^{2}=1-\frac{\sum_{i=1}^{n}\left(\phi_{\mathrm{True}}-\phi_{\text {Pred }}\right)^{2}}{\sum_{i=1}^{n}\left(\phi_{\mathrm{True}}-\bar{\phi}_{\mathrm{True}}\right)^{2}},
\label{eq16}
\end{split}
\end{equation}

\begin{equation}
\begin{split}
\mathrm{MSE}=<(\Delta \phi)^{2}>.
\label{eq17}
\end{split}
\end{equation}
Where $\phi_{True}$ is the true potential, $\phi_{Pred}$ is the predicted potential, and $\bar{\phi}_{True}$ is the mean true potential of a given test sample. In this study we used scatter $\sigma$, $R^{2}$ score and $MSE$ as our evaluating metrics.

\section{Result}\label{sec3}

In this paper, we predict the smooth potential of a point-charged particle in a spherical conductive container. First, we set the train and test set with 5000 and 1000 samples (\ref{subsec2}), then we train our models to predict smooth potential. We can calculate total potential by summating smooth and singular potential (more detailed in \ref{subsec1}). However, in this work, to compare our results with CAVIAR\cite{Biagooi2020}, we investigate the smooth potential.

\subsection{Random Forest}\label{subsec5}
\subsubsection{hyperparameter for RF}\label{subsub5}

\begin{figure}[!htbp]
\centering
\includegraphics[clip,width=\textwidth]{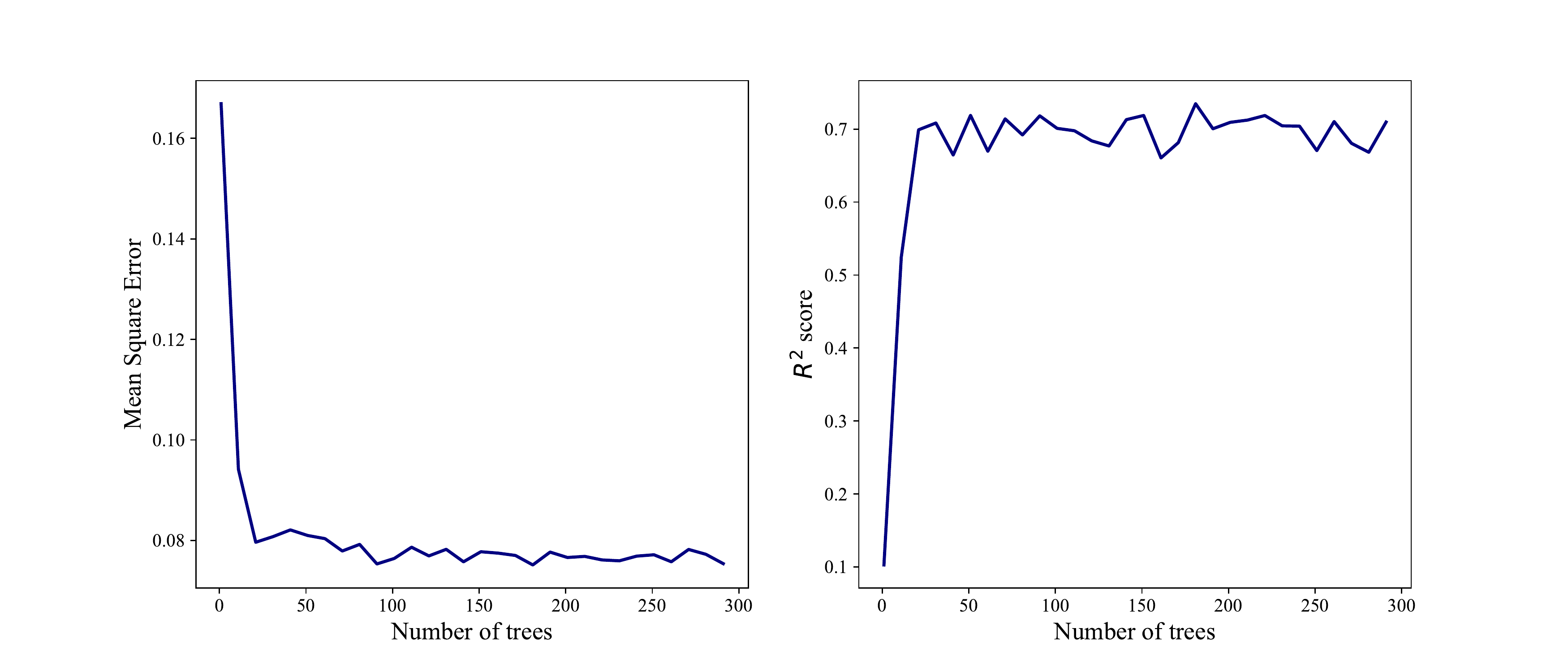}
\hrule
\caption{$MSE$  (left), and $R^{2}$ score (right) for 1000 different sample of Test set }\label{fig3}
\end{figure}

We optimize over the only hyperparameter, the number of trees in the forest that influences the fitting of the random forest model. In Fig.\ref{fig3}, we plot $MSE$ (the left panel) and $R^{2}$ score (the right one) as a function of the number of trees for the test set to determine the optimal hyperparameter, which we find 100 trees since progress after 100 trees are negligible. Afterward, we trained the RF model using 100 trees.

\FloatBarrier
\subsubsection{RF Prediction}\label{subsub6}
Fig.\ref{fig4} is illustrated the RF model with 100 trees. It shows that prediction is acceptable when the numeric value of potential is less than $0.3$ ($\phi_{True}<0.3$) while it could not predict precisely in the case of $\phi_{True}>=0.3$.
The graphs of Fig.\ref{fig4} compare the true potential $\phi_{True}$ and RF model predicted potential $\phi_{RF}$ for the train data set(left picture) and test data set(right image). The RF method is relatively fast; however, it works when the predicted potential is smooth and relatively small, it is not suitable in the case of point-charged particles (where a gradient of potential as well as its numeric value is high at the position of the charge). Furthermore, it fails to predict the potential near the boundaries since the gradient of the potential is considerable.

\begin{figure}[!htbp]
    \centering
    \subfigure[]{\includegraphics[width=0.48\textwidth]{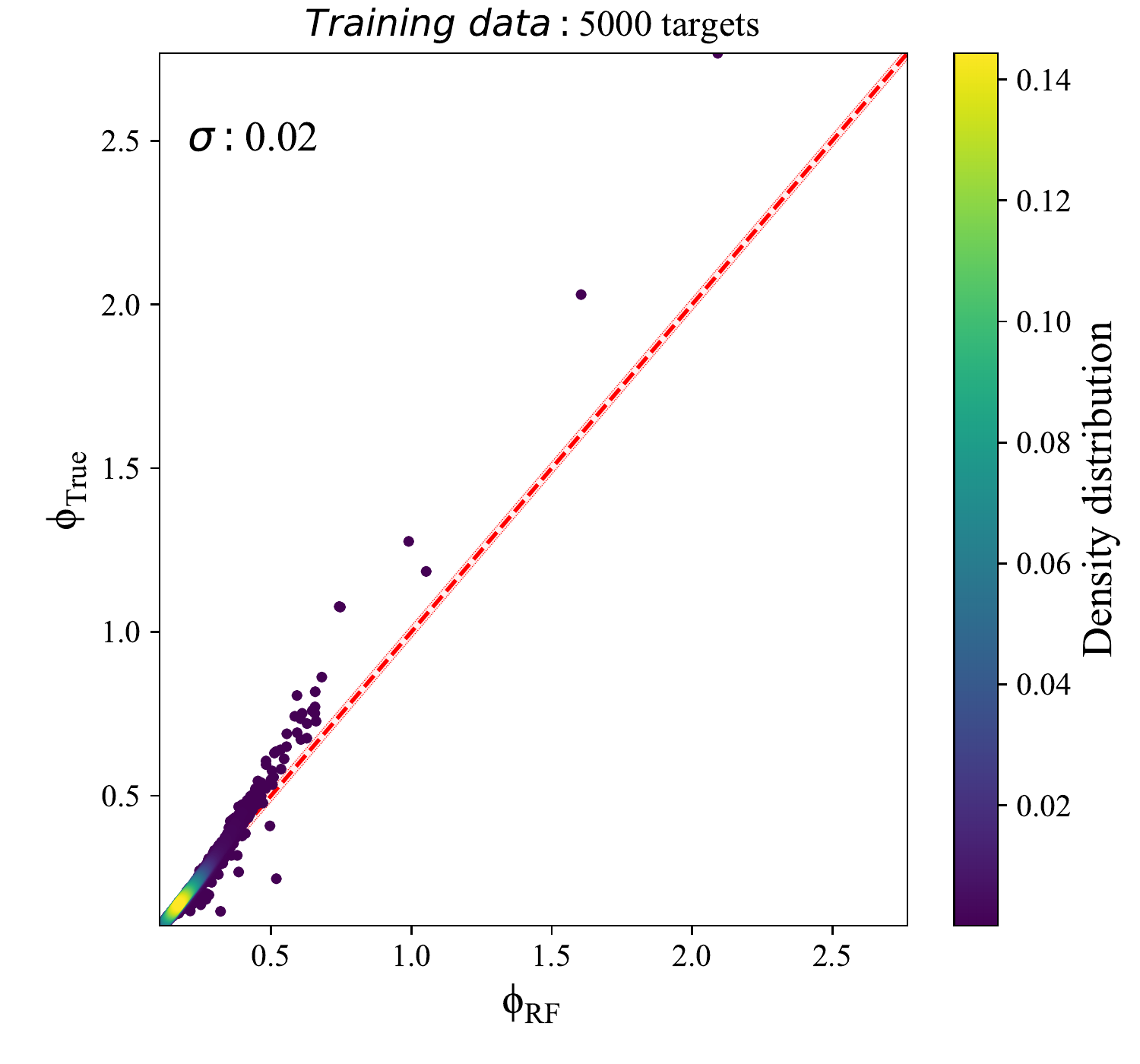}}
    \subfigure[]{\includegraphics[width=0.48\textwidth]{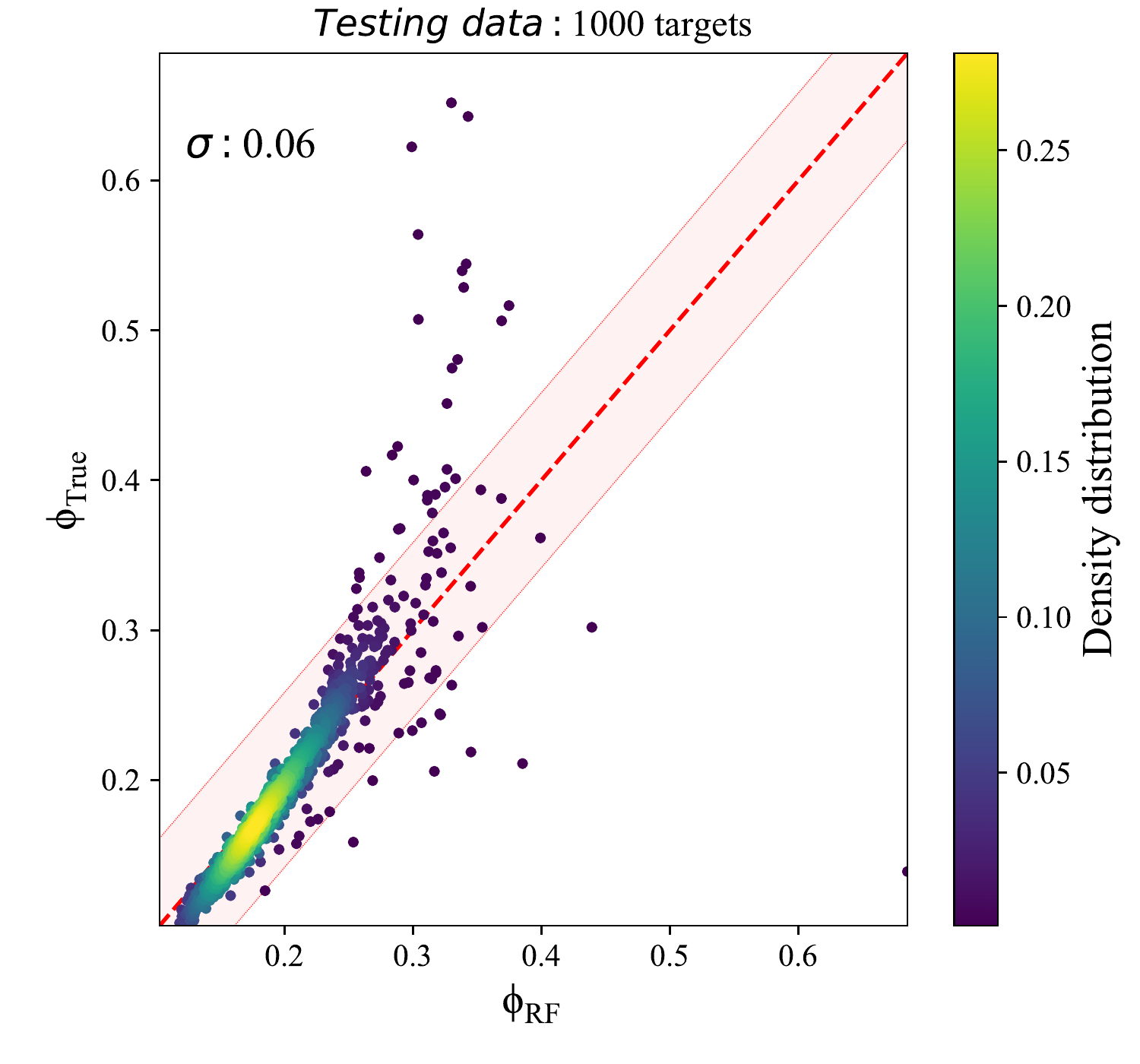}} 
    \caption{Potential estimation of RF model: a) show the train data set with 5000 samples with scatter of $\sigma=0.02$, b) show the test data set with 1000 samples with scatter of $\sigma=0.07$. The dashed red line shows where the predicted potential equals the true potential. The pink-shaded region marks $1\sigma$ scatter of
potential errors.  }
    \label{fig4}
\end{figure}
\FloatBarrier

\subsection{PINN based model and NN}\label{sub4}
By setting both $\lambda_{2}$ and $\lambda_{3}$ to zero in PINN-based model, one can get exactly NN model. Therefore we investigate
both models together and report their reulsts at the same time in the following section.
\subsubsection{hyperparameter for PINN and NN}\label{subsub7}

Unlike the RF model, we define several hyperparameters: a number of neurons, a number of layers, $\lambda_{2}$, and $\lambda_{3}$. To tune all hyperparameters, we train the model up to 100000 epochs, using the L-BFGS-B optimizer\cite{liu1989limited}, until the model's tolerance reaches the level of machine epsilon. For all layers except the last one, we use a $\tanh$ activation function. Table \ref{table:1} is reported the $MSE$ between the predicted and the same potential for a different value of hyperparameters; $\lambda_{2}=[0, 0.1, 0.2, 0.3]$, $\lambda_{3}=[0, 0.1, 0.2, 0.3, 0.4]$, number of hidden layers$=[1, 3, 5, 7]$ and number of neurons per hidden layer$=[10, 30, 50]$ for 1000 samples of the test set. As shown in Table  \ref{table:1}, we observe that a model with one hidden layer could not predict the potential well. Also, a model with ten neurons per layer could not work well. So, Table \ref{table:2} and Table \ref{table:3} reported the results for just $[3, 5, 7]$ layers as well as for $[30, 50]$ neurons per layer. We chose $\lambda_{4}= 0.0001$ to prevent over-fitting.  

\subsubsection{PINN and NN Prediction}\label{subsub5}
In Contrast with Table \ref{table:1}, Table \ref{table:2} is reported not only the $MSE$ but also the $R^{2} score$ of the test set. As can be seen in Table \ref{table:2}, the model with seven layers and 50 neurons per layer resulted better when $\lambda_{2}$ and $\lambda_{3}$ are $0.3,0.3$, or $0.2,0.4$, respectively. When $\lambda_{2}$ and $\lambda_{3}$ are zero, a standard neural network, the model has not worked well; it is observed from Tabel \ref{table:2} and Fig.\ref{fig3}.\\

\begin{figure}[!htbp]
    \centering
    
    \subfigure[]{\includegraphics[width=0.48\textwidth]{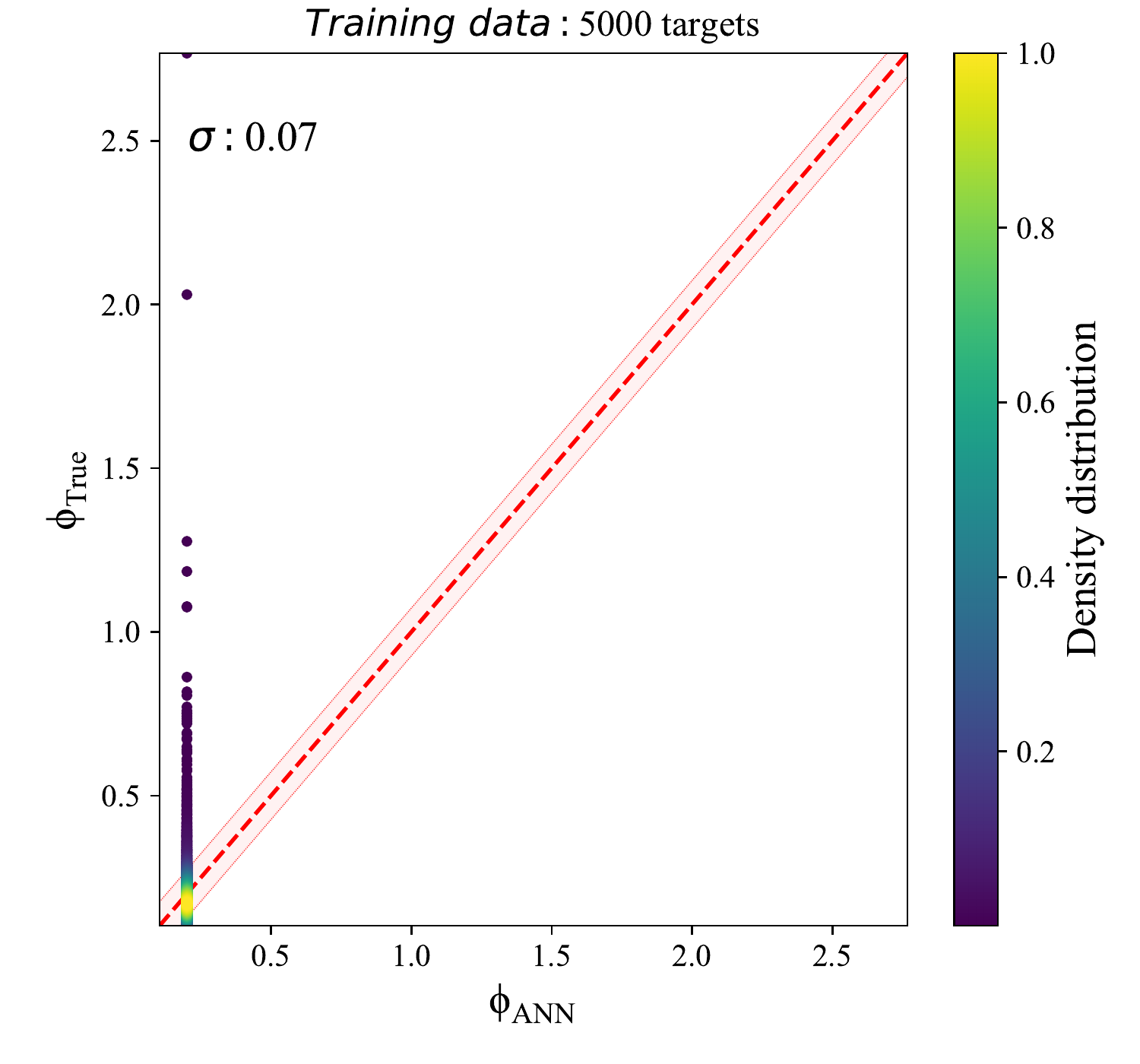}}
    \subfigure[]{\includegraphics[width=0.48\textwidth]{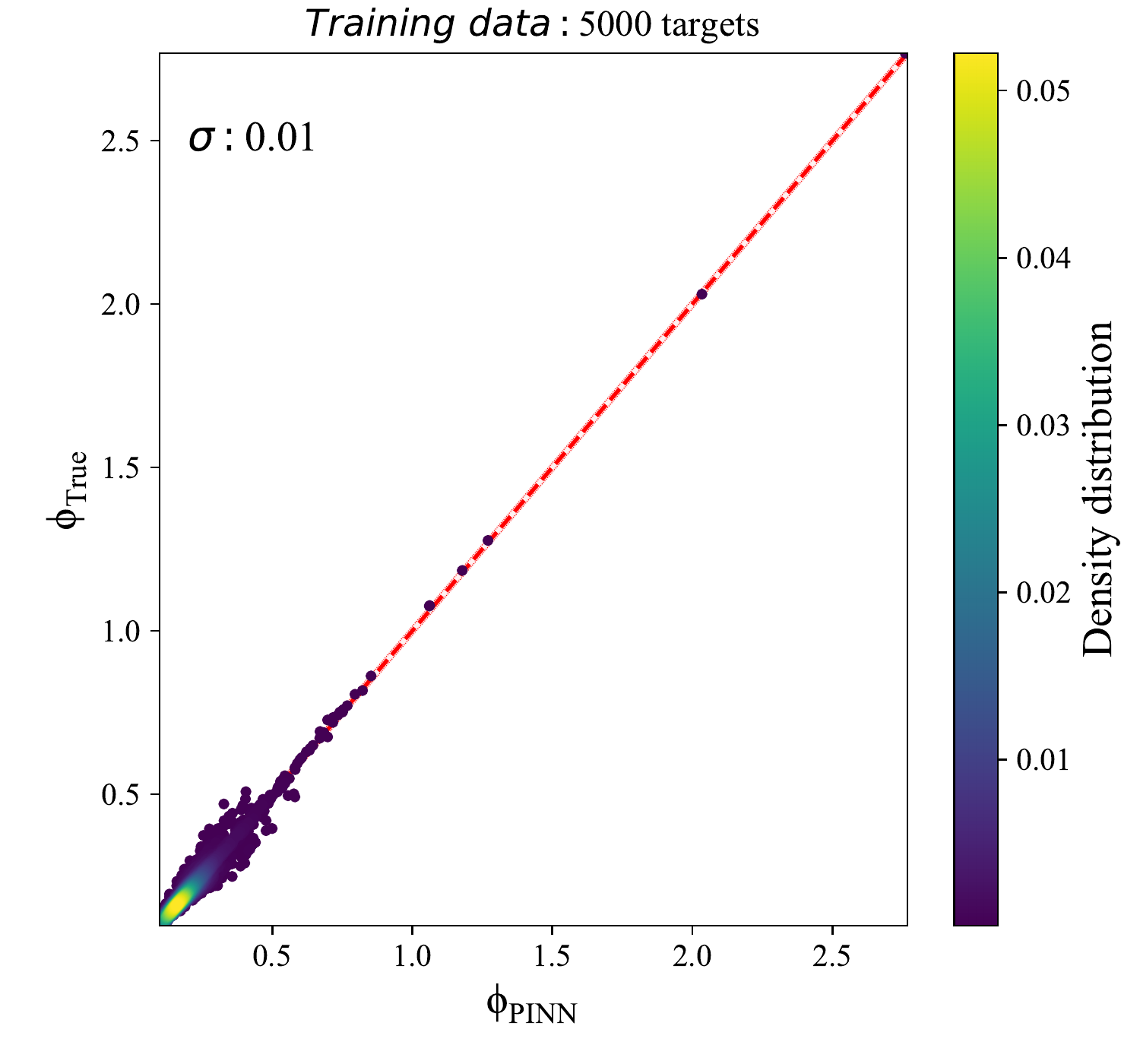}}
    
    \caption{Potential estimation of best-tuned (a) NN with scatter of $\sigma= 0.07$ and (b) PINN model on the 5000 samples of the train set with scatter of $\sigma= 0.01$. The dashed red line shows where the predicted potential equals the true potential. The pink-shaded region marks $1\sigma$ scatter of
potential errors. }
    \label{fig5}
\end{figure}
\begin{figure}[!htbp]
    \centering
    {\includegraphics[width=0.48\textwidth]{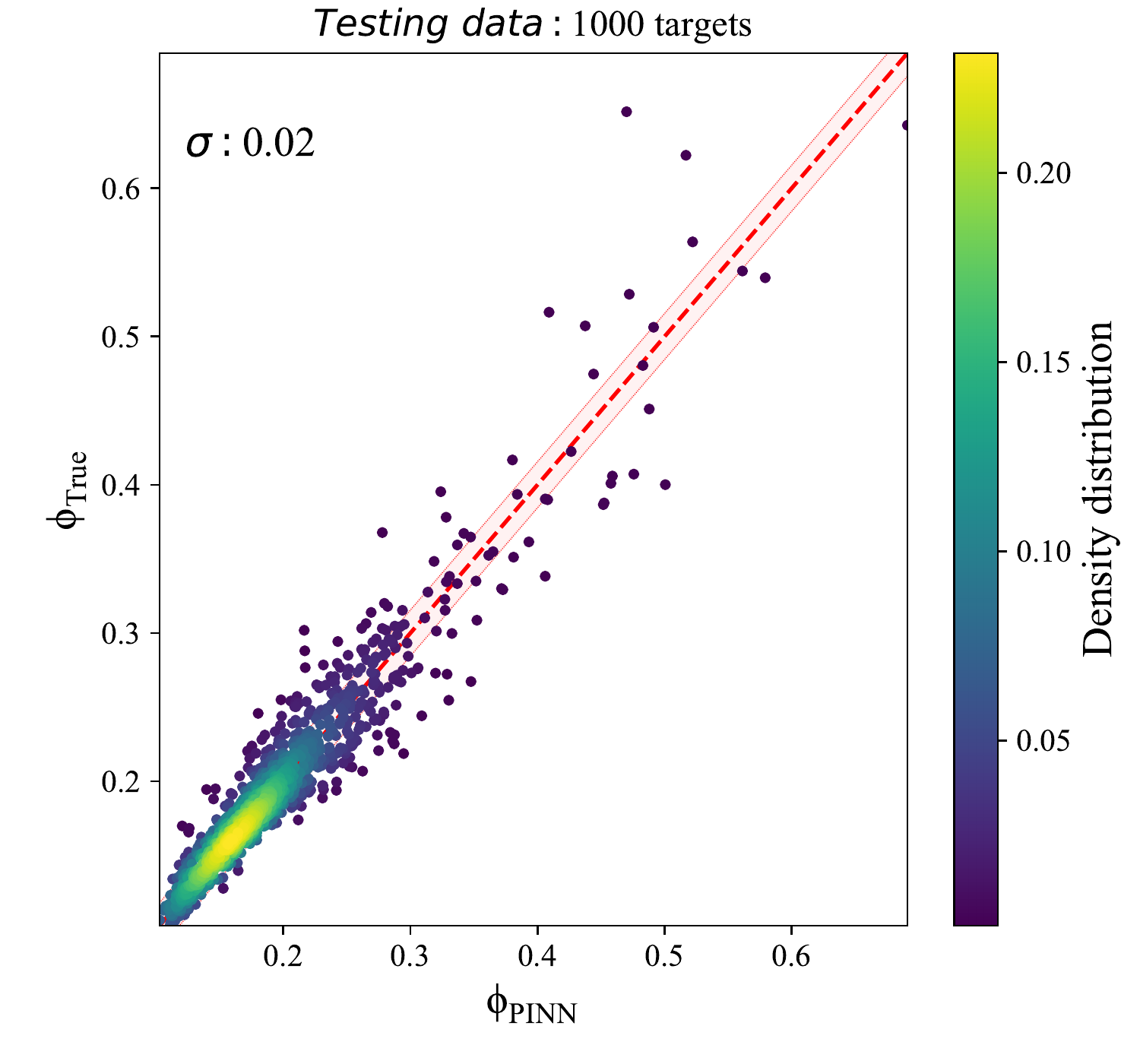}} 
    
    \caption{Potential estimation of best-tuned PINN model on the 1000 samples of the test set with scatter of $\sigma= 0.02$, $MSE= 0.069$ and $R^2 score= 0.851$. The dashed red line shows where the predicted potential equals the true potential. The pink-shaded region marks $1\sigma$ scatter of potential errors.}
    \label{fig6}
\end{figure}

\FloatBarrier
Both plots in Fig.\ref{fig5} compare the true and predicted potentials for the best-tuned NN and the best-tuned PINN model with 7 layers and 50 neurons per layer, $\lambda_{2}=0.3$ and $\lambda_{3}=0.3$- on the train set.
As can be seen, the NN model is not trained well, while the PINN-based model could predict the potential precisely with a scatter of 0.01. Although the PINN-based model predicts the train set well, aiming to clarify that over-fitting has not been accrued, we also evaluate the model on the test set, Fig.\ref{fig6}.

\FloatBarrier

\begin{center}
\begin{table}[!htbp]
    \centering
    \renewrobustcmd{\bfseries}{\fontseries{b}\selectfont}
    \renewrobustcmd{\boldmath}{}
    \renewcommand{\arraystretch}{1}
    \setlength{\tabcolsep}{4pt}
    \begin{tabular}{@{}l c||  *{2}{c} c| *{3}{c} @{}}
    \hline 
    \multicolumn{1}{c}{} &\multicolumn{1}{c}{} & \multicolumn{3}{c}{$MSE_{test}$}  & \multicolumn{3}{c}{$MSE_{test}$} \\
     \cmidrule(rl){3-5}  \cmidrule(rl){6-8}
    \multicolumn{1}{c}{$\lambda_{2}$} & \multicolumn{1}{c}{$\lambda_{3}$} & \multicolumn{1}{c}{\textbf{Neurons=10}} & \multicolumn{1}{c}{\textbf{Neurons=30}} & \multicolumn{1}{c}{\textbf{Neurons=50}} & \multicolumn{1}{c}{\textbf{Neurons=10}} & \multicolumn{1}{c}{\textbf{Neurons=30}} & \multicolumn{1}{c}{\textbf{Neurons=50} } \\
    \hline\hline
   
    & \multicolumn{1}{c}{}&\multicolumn{3}{c}{\textbf{Num of hidden layer:1}} &\multicolumn{3}{c}{\textbf{Num of hidden layer:5}} \\ \hline
    \multirow{5}{*}{0.0}  &0.0 &0.246 & 0.246 & 0.246 & 0.246 & 0.246 & 0.246 \\
    & 0.1 &0.162 & 0.145 & 0.124 & \bfseries0.099 & 0.145 &\bfseries 0.078 \\    
    & 0.2 &0.191 & 0.212 & 0.132 & 0.136 &\bfseries 0.062 &\bfseries 0.072 \\
    & 0.3 &0.175 & 0.159 & 0.171 & 0.14  &\bfseries 0.069 & 0.124 \\
    & 0.4 &0.24  & 0.171 & 0.18  & 0.191 &\bfseries 0.064 &\bfseries 0.076 \\ \hline
    \multirow{5}{*}{0.1}  &0.0 &0.246 & 0.247 & 0.247 & 0.246 & 0.246 & 0.246 \\
    & 0.1 &0.222 & 0.143 & 0.117 & 0.107 & 0.245 & 0.192 \\
    & 0.2 &0.201 & 0.165 & 0.129 & 0.243 & \bfseries0.079 & \bfseries0.062 \\
    & 0.3 &0.22  & 0.159 & 0.16  & 0.112 & \bfseries0.094 &\bfseries 0.071 \\
    & 0.4 &0.249 & 0.177 & 0.243 & 0.205 & \bfseries0.088 & \bfseries0.07  \\\hline
    \multirow{5}{*}{0.2}  &0.0 &0.246 & 0.246 & 0.246 & 0.246 & 0.246 & 0.246 \\
    & 0.1 &0.158 & 0.144 & 0.154 & 0.235 & 0.161 & \bfseries 0.083 \\
    & 0.2 &0.225 & 0.174 & 0.174 & 0.198 &\bfseries 0.097 & \bfseries0.071 \\
    & 0.3 &0.223 & 0.143 & 0.158 & 0.192 & \bfseries0.081 & \bfseries0.075 \\
    & 0.4 &0.197 & 0.185 & 0.189 & 0.216 &\bfseries 0.065 & 0.118 \\\hline
    \multirow{5}{*}{0.3}  &0.0 &0.246 & 0.246 & 0.246 & 0.246 & 0.246 & 0.246 \\
    & 0.1 &0.21  & 0.132 & 0.129 & 0.245 & \bfseries0.088 & \bfseries0.098 \\
    & 0.2 &0.199 & 0.139 & 0.157 & 0.209 & 0.117 & \bfseries0.065 \\
    & 0.3 &0.216 & 0.241 & 0.202 & 0.182 & 0.168 & \bfseries0.08  \\
    & 0.4 &0.225 & 0.182 & 0.196 & 0.239 & \bfseries0.083 & \bfseries0.097 \\\hline
    & \multicolumn{1}{c}{}&\multicolumn{3}{c}{\textbf{Num of hidden layer:3}} &\multicolumn{3}{c}{\textbf{Num of hidden layer:7}} \\ \hline
    \multirow{5}{*}{0.0}  &0.0 &0.246 & 0.246 & 0.246 & 0.246 & 0.246 & 0.246 \\
    & 0.1 &0.157 & 0.104 & \bfseries0.086 & 0.244 & 0.237 & 0.247 \\
    & 0.2 &0.151 & \bfseries0.088 & \bfseries0.086 & 0.238 & 0.087 & 0.105 \\
    & 0.3 &0.13  & \bfseries0.078 & \bfseries0.089 & 0.244 & \bfseries0.07  & \bfseries0.067 \\
    & 0.4 &0.202 & 0.173 & 0.102 & 0.188 & \bfseries0.068 & \bfseries0.063 \\\hline
    \multirow{5}{*}{0.1}  &0.0 &0.246 & 0.246 & 0.246 & 0.246 & 0.246 & 0.246 \\
    & 0.1 &\bfseries0.099 & \bfseries0.089 & 0.132 & 0.244 & 0.244 & 0.244 \\
    & 0.2 &0.198 & \bfseries0.093 & \bfseries0.077 & 0.23  & 0.221 & \bfseries0.069 \\
    & 0.3 &0.188 & \bfseries0.088 & \bfseries0.083 & 0.224 & 0.223 & \bfseries0.082 \\
    & 0.4 &0.262 & 0.104 & \bfseries0.079 & 0.213 & \bfseries0.072 & \bfseries\bfseries0.074 \\\hline
    \multirow{5}{*}{0.2}  &0.0 &0.246 & 0.246 & 0.246 & 0.246 & 0.246 & 0.246 \\
    & 0.1 &0.119 & 0.146 & \bfseries0.082 & 0.244 & 0.244 & 0.245 \\
    & 0.2 &0.172 & 0.09  & \bfseries0.086 & 0.244 & 0.244 &\bfseries 0.071 \\
    & 0.3 &0.179 & 0.084 & \bfseries0.085 & 0.225 & 0.251 & \bfseries0.08  \\
    & 0.4 &0.197 & 0.185 & \bfseries0.086 & 0.256 & 0.166 & \bfseries0.067 \\\hline
    \multirow{5}{*}{0.3}  &0.0 &0.246 & 0.246 & 0.246 & 0.246 & 0.246 & 0.246 \\
    & 0.1 &0.17  & 0.125 & \bfseries0.082 & 0.244 & 0.244 & 0.244 \\
    & 0.2 &0.199 & 0.117 & \bfseries0.096 & 0.244 & \bfseries0.083 & \bfseries0.08  \\
    & 0.3 &0.224 & 0.107 & \bfseries0.088 & 0.202 & 0.235 & \bfseries0.069 \\
    & 0.4 &0.199 & 0.222 & 0.152 & 0.236 & \bfseries0.097 & \bfseries0.075 \\ \hline

    \end{tabular}
    \vspace{0.25cm}
    \hrule
    \vspace{0.25cm}
    \caption{$ MSE $ between the predicted and the exact potential $\phi(x)$ for a different value of $\lambda_{2}$, and $\lambda_{3}$, and the different number of hidden layers, and neurons per hidden layer in PINN for 1000 different sample of the Test set. Here, $\lambda_{4}=0.0001$ is fixed and $\lambda_{1}= 1- ( \lambda_{2} + \lambda_{3} + \lambda_{4})$. In this table, the bold number means $MSE<0.1$.}
    \label{table:1}
    \end{table}
\end{center}

\begin{center}
\begin{table}[!htbp]
    \centering
    \renewrobustcmd{\bfseries}{\fontseries{b}\selectfont}
    \renewrobustcmd{\boldmath}{}
    \renewcommand{\arraystretch}{1}
    \setlength{\tabcolsep}{12pt}
    \begin{tabular}{@{}l c|| c c c c c c @{}} 
    \hline 
    \multicolumn{2}{c}{}  &\multicolumn{2}{c}{{$Num_{layer}$=3}} & \multicolumn{2}{c}{{$Num_{layer}$=5}} & \multicolumn{2}{c}{{$Num_{layer}$=7}}\\
    \cmidrule(rl){3-4} \cmidrule(rl){5-6} \cmidrule(rl){7-8}  
    {$\lambda_{2}$} & {$\lambda_{3}$}  & $MSE$ & $R^{2}$ & $MSE$ & $R^{2}$  & $MSE$ & $R^{2}$\\
    \hline\hline
    \multicolumn{2}{c}{}&\multicolumn{2}{c}{}&\multicolumn{2}{c}{{$Num_{neuron}$=30}} &\multicolumn{2}{c}{} \\ \hline
   
    \multirow{5}{*}{0.0}  &0.0 &0.246 & 0     & 0.246 & 0     & 0.246 & 0      \\
    & 0.1 &0.104 & 0.774 & 0.145 & 0.365 & 0.237 & -14.7  \\
    & 0.2 &0.088 & 0.85  & 0.062 & 0.889 & 0.087 & 0.807  \\
    & 0.3 &0.078 & 0.826 & 0.069 & 0.884 & \bfseries0.07  &\bfseries 0.904  \\
    & 0.4 &0.173 & 0.551 & 0.064 & 0.863 & 0.068 & 0.868  \\ \hline
    \multirow{5}{*}{0.1}  &0.0 &0.246 & 0     & 0.246 & 0     & 0.246 & 0      \\
    & 0.1 &0.089 & 0.796 & 0.245 & -1478 & 0.244 & 0      \\
    & 0.2 &0.093 & 0.825 & 0.079 & 0.852 & 0.221 & -2.40 \\
    & 0.3 &0.088 & 0.793 & 0.094 & 0.714 & 0.223 & 0.094  \\
    & 0.4 &0.104 & 0.719 & 0.088 & 0.793 & 0.072 & 0.887  \\ \hline
    \multirow{5}{*}{0.2}  &0.0 &0.246 & 0     & 0.246 & 0     & 0.246 & 0      \\
    & 0.1 &0.146 & 0.64  & 0.161 & 0.566 & 0.244 & 0      \\
    & 0.2 &0.09  & 0.769 & 0.097 & 0.75  & 0.244 & -3167  \\
    & 0.3 &0.084 & 0.819 & 0.081 & 0.857 & 0.251 & -2.32 \\
    & 0.4 &0.185 & 0.521 & 0.065 & 0.874 & 0.166 & 0.07   \\ \hline
    \multirow{5}{*}{0.3}  &0.0 &0.246 & 0     & 0.246 & 0     & 0.246 & 0      \\
    & 0.1 &0.125 & 0.626 & 0.088 & 0.747 & 0.244 & 0      \\
    & 0.2 &0.117 & 0.666 & 0.117 & 0.689 & 0.083 & 0.863  \\
    & 0.3 &0.107 & 0.67  & 0.168 & 0.59  & 0.235 & 0 \\
    & 0.4 &0.222 & 0.25  & 0.083 & 0.837 & 0.097 & 0.804  \\ \hline
    \multicolumn{2}{c}{}&\multicolumn{2}{c}{}&\multicolumn{2}{c}{{$Num_{neuron}$=50}} &\multicolumn{2}{c}{} \\ \hline
    \multirow{5}{*}{0.0}  &0.0 &0.246 & 0     & 0.246 & 0     & 0.246 & 0      \\
    & 0.1 &0.086 & 0.832 & 0.078 & 0.865 & 0.247 & -254   \\
    & 0.2 &0.086 & 0.838 & 0.072 & 0.873 & 0.105 & 0.711  \\
    & 0.3 &0.089 & 0.767 & 0.124 & 0.66  & 0.067 & 0.875  \\
    & 0.4 &0.102 & 0.734 & 0.076 & 0.846 & 0.063 & 0.849  \\ \hline
    \multirow{5}{*}{0.1}  &0.0 &0.246 & 0     & 0.246 & 0     & 0.246 & 0      \\
    & 0.1 &0.132 & 0.64  & 0.192 & -0.48 & 0.244 & 0      \\
    & 0.2 &0.077 & 0.85  & 0.062 & 0.888 & 0.069 & 0.897  \\
    & 0.3 &0.083 & 0.837 & 0.071 & 0.895 & 0.082 & 0.774  \\
    & 0.4 &0.079 & 0.847 & 0.07  & 0.887 & 0.074 & 0.863  \\ \hline
    \multirow{5}{*}{0.2}  &0.0 &0.246 & 0     & 0.246 & 0     & 0.246 & 0      \\
    & 0.1 &0.082 & 0.756 & 0.083 & 0.859 & 0.245 & -1774  \\
    & 0.2 &0.086 & 0.796 & 0.071 & 0.88  & \bfseries0.071 &\bfseries 0.908  \\
    & 0.3 &0.085 & 0.799 & 0.075 & 0.888 & 0.08  & 0.87   \\
    & 0.4 &0.086 & 0.8   & 0.118 & 0.727 & \bfseries0.067 & \bfseries0.902  \\ \hline
    \multirow{5}{*}{0.3}  &0.0 &0.246 & 0     & 0.246 & 0     & 0.246 & 0      \\
    & 0.1 &0.082 & 0.75  & 0.098 & 0.753 & 0.244 & 0      \\
    & 0.2 &0.096 & 0.777 & 0.065 & 0.894 & 0.08  & 0.798  \\
    & 0.3 &0.088 & 0.785 & 0.08  & 0.834 & \bfseries0.069 & \bfseries0.902  \\
    & 0.4 &0.152 & 0.557 & 0.097 & 0.818 & 0.075 & 0.867 \\ \hline
   
    \end{tabular}
    \vspace{0.25cm}
    \hrule
    \vspace{0.25cm}
    \caption{$ MSE $, and $R^{2}$ score between the predicted and the exact potential $\phi(x)$ for a different value of $\lambda_{2}$, and $\lambda_{3}$, and the different number of hidden layers, and neurons per hidden layer in PINN for 1000 different samples of the Test set. Here, $\lambda_{4}=0.0001$ is fixed and $\lambda_{1}= 1- ( \lambda_{2} + \lambda_{3} + \lambda_{4})$. In this table bold numbers show cases with $MSE<0.1$ and $R^{2} score> 0.9$.}
    \label{table:2}
    \end{table}
\end{center}

\begin{center}
\begin{table}[!htbp]
    \centering
    \renewrobustcmd{\bfseries}{\fontseries{b}\selectfont}
    \renewrobustcmd{\boldmath}{}
    \renewcommand{\arraystretch}{1}
    \setlength{\tabcolsep}{12pt}
    \begin{tabular}{@{}l c|| c c c c c c @{}} 
    \hline 
    \multicolumn{2}{c}{}  &\multicolumn{2}{c}{{$Num_{layer}$=3}} & \multicolumn{2}{c}{{$Num_{layer}$=5}} & \multicolumn{2}{c}{{$Num_{layer}$=7}}\\
    \cmidrule(rl){3-4} \cmidrule(rl){5-6} \cmidrule(rl){7-8}  
    {$\lambda_{2}$} & {$\lambda_{3}$}  & $MSE$ & $R^{2}$ & $MSE$ & $R^{2}$  & $MSE$ & $R^{2}$\\
    \hline\hline
    \multicolumn{2}{c}{}&\multicolumn{2}{c}{}&\multicolumn{2}{c}{{$Num_{neuron}$=30}} &\multicolumn{2}{c}{} \\ \hline
   
    \multirow{5}{*}{0.0}  &0.0& 0.281 & 0     & 0.281 & 0      & 0.281 & 0      \\
    & 0.1 &0.14  & 0.71  & 0.148 & 0.364  & 0.275 & -15.28 \\
    & 0.2 &0.118 & 0.764 & 0.077 & 0.847  & 0.126 & 0.646  \\
    & 0.3 &0.104 & 0.737 & 0.09  & 0.782  & 0.089 & 0.835  \\
    & 0.4 &0.208 & 0.587 & 0.072 & 0.842  & 0.083 & 0.822  \\\hline
	\multirow{5}{*}{0.1}  &0.0&0.281 & 0     & 0.281 & 0      & 0.281 & 0      \\
    & 0.1 &0.105 & 0.808 & 0.28  & -1801  & 0.28  & 0      \\
    & 0.2 &0.112 & 0.775 & 0.09  & 0.845  & 0.259 & -2.137 \\
    & 0.3 &0.105 & 0.793 & 0.123 & 0.697  & 0.264 & 0.07   \\
    & 0.4 &0.116 & 0.735 & 0.12  & 0.684  & 0.092 & 0.813  \\\hline
	\multirow{5}{*}{0.2}  &0.0&0.281 & 0     & 0.281 & 0      & 0.281 & 0      \\
    & 0.1 &0.183 & 0.601 & 0.201 & 0.484  & 0.28  & 0      \\
    & 0.2 &0.118 & 0.739 & 0.114 & 0.68   & 0.28  & -3620  \\
    & 0.3 &0.095 & 0.829 & 0.109 & 0.799  & 0.275 & -1.82  \\
    & 0.4 &0.209 & 0.586 & 0.078 & 0.851  & 0.205 & 0.183  \\\hline
	\multirow{5}{*}{0.3}  &0.0&0.281 & 0     & 0.281 & 0      & 0.281 & 0      \\
    & 0.1 &0.154 & 0.663 & 0.105 & 0.743  & 0.28  & 0      \\
    & 0.2 &0.13  & 0.74  & 0.139 & 0.688  & 0.109 & 0.725  \\
    & 0.3 &0.143 & 0.565 & 0.2   & 0.604  & 0.271 & 0.012  \\
    & 0.4 &0.269 & 0.356 & 0.158 & 0.364  & 0.118 & 0.726  \\\hline
	\multicolumn{2}{c}{}&\multicolumn{2}{c}{}&\multicolumn{2}{c}{{$Num_{neuron}$=50}} &\multicolumn{2}{c}{} \\ \hline
	\multirow{5}{*}{0.0}  &0.0&0.281 & 0     & 0.281 & 0      & 0.281 & 0      \\
    & 0.1 &0.116 & 0.764 & 0.112 & 0.769  & 0.28  & -310   \\
    & 0.2 &0.1   & 0.837 & 0.092 & 0.829  & 0.135 & 0.598  \\
    & 0.3 &0.101 & 0.79  & 0.145 & 0.719  & 0.083 & 0.843  \\
    & 0.4 &0.117 & 0.706 & 0.096 & 0.801  & 0.106 & 0.662  \\\hline
	\multirow{5}{*}{0.1}  &0.0&0.281 & 0     & 0.281 & 0      & 0.281 & 0      \\
    & 0.1 &0.149 & 0.657 & 0.227 & -0.465 & 0.28  & 0      \\
    & 0.2 &0.121 & 0.658 & 0.077 & 0.832  & 0.09  & 0.824  \\
    & 0.3 &0.108 & 0.719 & 0.096 & 0.817  & 0.102 & 0.734  \\
    & 0.4 &0.117 & 0.716 & 0.085 & 0.843  & 0.093 & 0.842  \\\hline
	\multirow{5}{*}{0.2}  &0.0&0.281 & 0     & 0.281 & 0      & 0.281 & 0      \\
    & 0.1 &0.094 & 0.782 & 0.112 & 0.777  & 0.28  & -2164  \\
    & 0.2 &0.114 & 0.758 & 0.094 & 0.785  & 0.091 & 0.854  \\
    & 0.3 &0.101 & 0.762 & 0.093 & 0.818  & 0.104 & 0.771  \\
    & 0.4 &0.124 & 0.644 & 0.143 & 0.703  & 0.088 & 0.849  \\\hline
	\multirow{5}{*}{0.3}  &0.0&0.281 & 0     & 0.281 & 0      & 0.281 & 0      \\
    & 0.1 &0.097 & 0.772 & 0.115 & 0.727  & 0.28  & 0      \\
    & 0.2 &0.112 & 0.768 & 0.096 & 0.807  & 0.167 & 0.363  \\
    & 0.3 &0.11  & 0.777 & 0.113 & 0.682  &\bfseries 0.089 &\bfseries 0.851  \\
    & 0.4 &0.177 & 0.563 & 0.114 & 0.76   & 0.103 & 0.741  \\\hline
    \end{tabular}
    \vspace{0.25cm}
    \hrule
    \vspace{0.25cm}
    \caption{$ MSE $ and $R^{2}$ score between the predicted and the exact potential $\phi(x)$ for a different value of $\lambda_{2}$, and $\lambda_{3}$, and the different number of hidden layers, and neurons per hidden layer in PINN for 1000 different samples of the Extrapolate set. Here, $\lambda_{4}=0.0001$ is fixed and $\lambda_{1}= 1- ( \lambda_{2} + \lambda_{3} + \lambda_{4})$. In this table bold number shows the best hyperparameters for our PINN-based model.}
    \label{table:3}
    \end{table}
\end{center}
\FloatBarrier

\subsection{Comparison }\label{sec4}
\begin{figure}[!htbp]
    \centering 
    \subfigure[]{\includegraphics[width=0.45\textwidth]{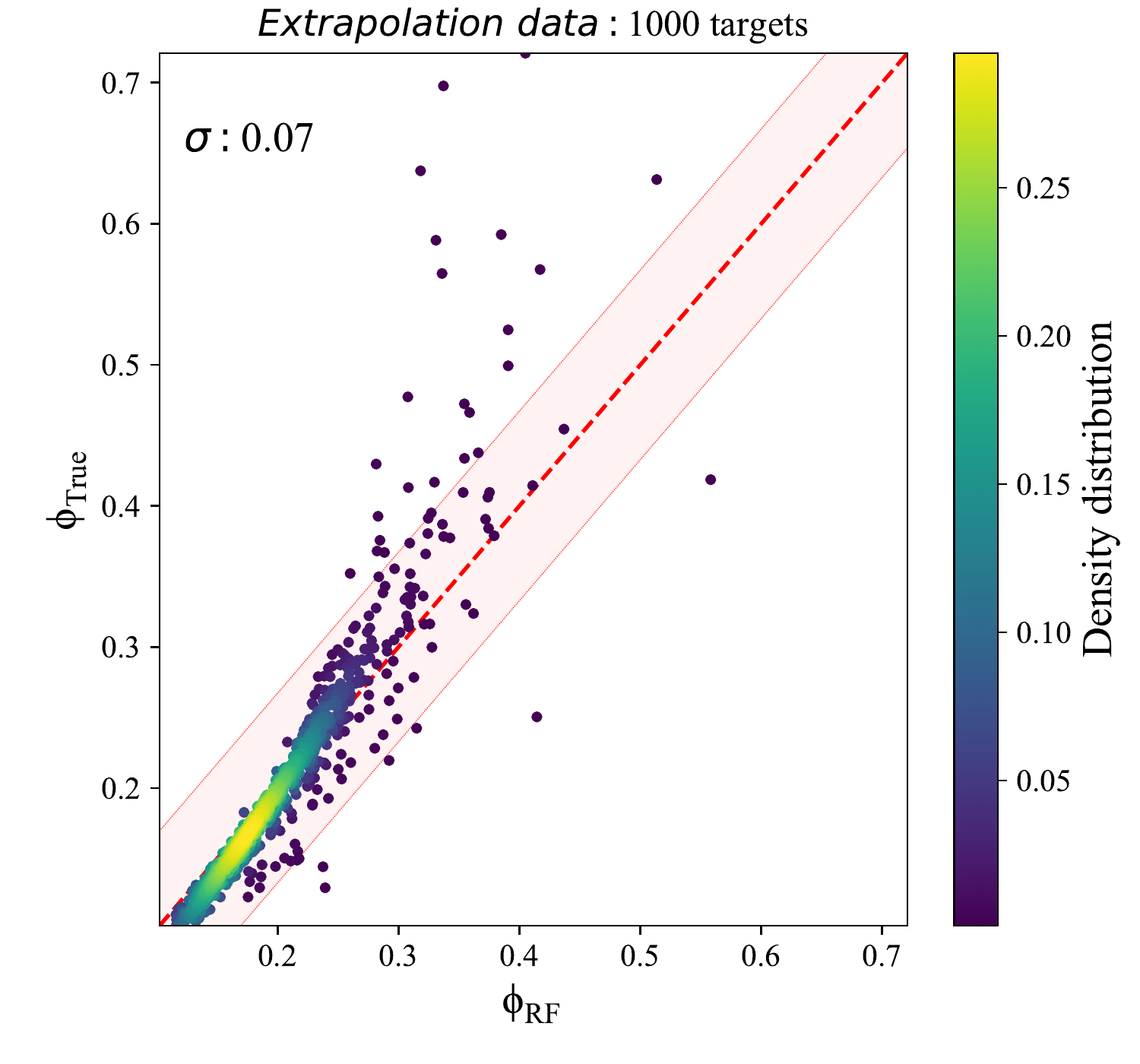}}
    \subfigure[]{\includegraphics[width=0.45\textwidth]{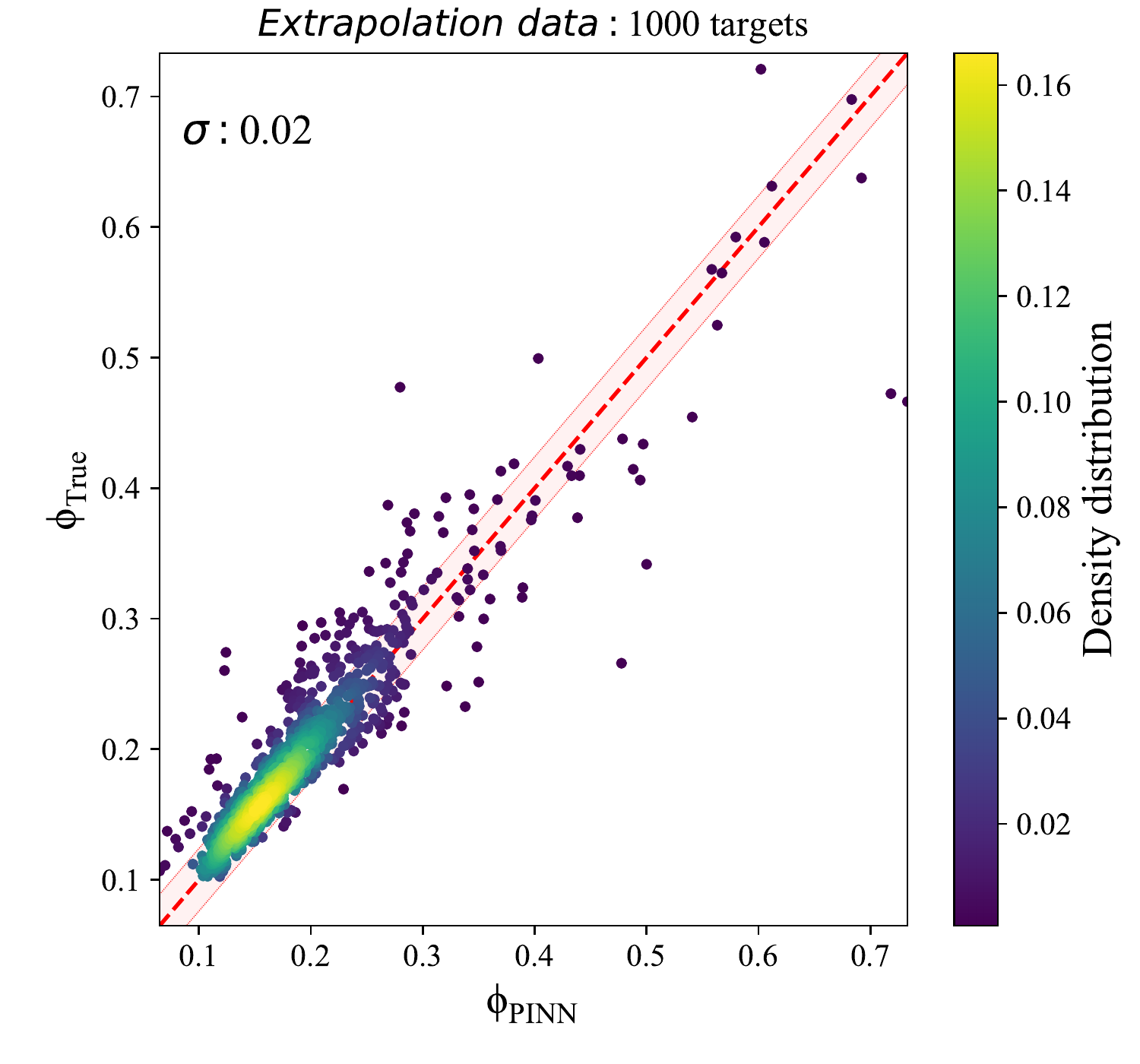}}
    \caption{Potential estimation of best-tuned (a) RF with scatter of $\sigma= 0.07$, and (b) PINN-based model with scatter of $\sigma= 0.02$ on the 1000 samples of the Extrapolation set. The dashed red line shows where the predicted potential equals the true potential. The pink-shaded region marks $1\sigma$ scatter of
potential errors.}
    \label{fig7}
\end{figure}

We evaluate RF, NN, and PINN to estimate the potential of point-charged particles surrounded by conductive walls. According to Fig.\ref{fig5}, NN was not trained well, while RF and PINN-based models could predict potential precisely. However, RF did not work well to estimate $\phi_{True}>0.3$. Apart from this, the best model could estimate not only the potential of the train and the test sets but also the potential of point-charged particles that are not in the train or test set. So we evaluate the best tuned-PINN model and RF on the extrapolation samples; the results are reported in Table \ref{table:3}. As can be seen in Fig.\ref{fig7} PINN-based model could predict the potential of newly charged particles better than the RF model, where PINN could predict $\phi_{True}>0.3$ by far better than RF. 

\FloatBarrier

\subsection{Generalization (Multi charged particles)}\label{sec5}
\begin{figure}[!htbp]
\centering
\includegraphics[scale=0.5]{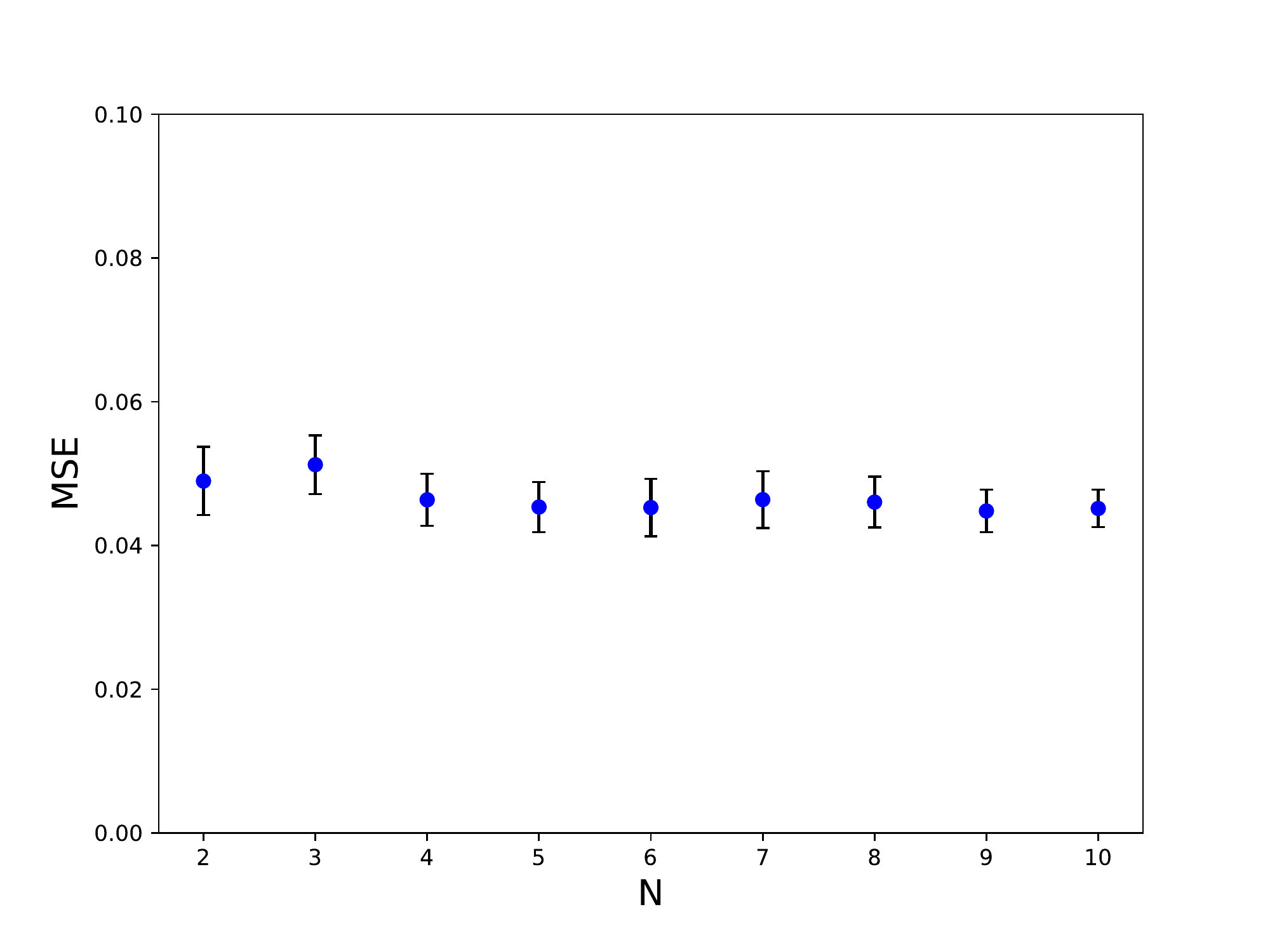}
\hrule
\caption{$MSE$ between true and predicted potential as a function of charged particles number; in 100 different problems}\label{fig8}
\end{figure}

For generalization, we test the PINN-based model with $\lambda_{2}=0.3$ and $\lambda_{3}=0.3$ for the case of more than one charged particle surrounded with conductive boundaries. Since the Laplace equation is a linear function, we predict the potential of each charged particle and then calculate the total potential with a superposition of the corresponding predicted potential. After that, we report the $MSE$ between the predicted smooth potential and the exact smooth solution, which is calculated by the image charges method. Fig.\ref{fig8} shows the relation between $MSE$ and $N$, the number of charged particles. As expected, the $MSE$ is independent of the number of charged particles. Therefore, 
it leads to the fact that we can also use this method for problems with any desired particles.

\FloatBarrier

\section{Conclusion}

In this study, we have trained a machine to predict the smooth potential of charged components surrounded by conductive boundaries. In this scene, the total potential could be easily calculated by the summation of predicted smooth potential with singular potential due to the PLT algorithm. The reference set consists of analytic solutions, the solution of the image charge method, which is split into a train set with 5000 samples and a test set with 1000 samples. To check the accuracy of our model, we set another data set called extrapolation set consisting of 1000 samples with different boundary conditions which were not in the train or even test set. Our main conclusion can be summarized as follows:
\begin{itemize}
\item We find that the PINN-based model trained better than RF and NN models. RF could not predict high potential; on the other hand, the NN could not be trained well at all.
\item our PINN-based model could predict the potential of the test set with $MSE=0.069$, $R^{2} score=0.902$, and scatter $\sigma=0.02$. It also could predict the potential of the extrapolation set with $MSE=0.089$, $R^{2} score=0.851$, and scatter $\sigma=0.02$.
\item Since the Laplace equation is a linear equation, the trained model could predict the potential of more than one charged particle by summating every particle's predicted potential. Besides, we show that the $MSE$ of more than one particle is independent of a number of particles.  

\end{itemize}

\bibliographystyle{unsrtnat}
\bibliography{references}  

\begin{thebibliography}{25}
\providecommand{\natexlab}[1]{#1}
\providecommand{\url}[1]{\texttt{#1}}
\expandafter\ifx\csname urlstyle\endcsname\relax
  \providecommand{\doi}[1]{doi: #1}\else
  \providecommand{\doi}{doi: \begingroup \urlstyle{rm}\Url}\fi

\bibitem[Miller and Simon(2008)]{miller2008electrochemical}
John~R Miller and Patrice Simon.
\newblock Electrochemical capacitors for energy management.
\newblock \emph{science}, 321\penalty0 (5889):\penalty0 651--652, 2008.

\bibitem[Salanne et~al.(2016)Salanne, Rotenberg, Naoi, Kaneko, Taberna, Grey,
  Dunn, and Simon]{salanne2016efficient}
Mathieu Salanne, Benjamin Rotenberg, Katsuhiko Naoi, Katsumi Kaneko, P-L
  Taberna, Clare~P Grey, Bruce Dunn, and Patrice Simon.
\newblock Efficient storage mechanisms for building better supercapacitors.
\newblock \emph{Nature Energy}, 1\penalty0 (6):\penalty0 1--10, 2016.

\bibitem[Simon and Gogotsi(2008)]{Simon2008}
Patrice Simon and Yury Gogotsi.
\newblock Materials for electrochemical capacitors, 2008.
\newblock ISSN 14761122.

\bibitem[Jackson(1962)]{Jackson1962}
John~David Jackson.
\newblock Jackson - classical electrodynamics, 1962.
\newblock ISSN 00029505.

\bibitem[Jin(2014)]{Jin2014}
Jian-Ming Jin.
\newblock The finite element method in electromagnetics, 3rd edition.
\newblock \emph{Journal of Chemical Information and Modeling}, 2014.

\bibitem[S. et~al.(1991)S., Golub, and Loan]{GeneH}
G.~W. S., Gene~H. Golub, and Charles F.~Van Loan.
\newblock Matrix computations.
\newblock \emph{Mathematics of Computation}, 56, 1991.
\newblock ISSN 00255718.
\newblock \doi{10.2307/2008552}.

\bibitem[Tyagi et~al.(2007)Tyagi, Arnold, and Holm]{Tyagi2007}
Sandeep Tyagi, Axel Arnold, and Christian Holm.
\newblock Icmmm2d: An accurate method to include planar dielectric interfaces
  via image charge summation.
\newblock \emph{Journal of Chemical Physics}, 127, 2007.
\newblock ISSN 00219606.
\newblock \doi{10.1063/1.2790428}.

\bibitem[Tyagi et~al.(2008)Tyagi, Arnold, and Holm]{Tyagi2008}
Sandeep Tyagi, Axel Arnold, and Christian Holm.
\newblock Electrostatic layer correction with image charges: A linear scaling
  method to treat slab 2d+h systems with dielectric interfaces.
\newblock \emph{Journal of Chemical Physics}, 129, 2008.
\newblock ISSN 00219606.
\newblock \doi{10.1063/1.3021064}.

\bibitem[Tyagi et~al.(2010)Tyagi, Süzen, Sega, Barbosa, Kantorovich, and
  Holm]{Tyagi2010}
Sandeep Tyagi, Mehmet Süzen, Marcello Sega, Marcia Barbosa, Sofia~S.
  Kantorovich, and Christian Holm.
\newblock An iterative, fast, linear-scaling method for computing induced
  charges on arbitrary dielectric boundaries.
\newblock \emph{Journal of Chemical Physics}, 132, 2010.
\newblock ISSN 00219606.
\newblock \doi{10.1063/1.3376011}.

\bibitem[Kesselheim et~al.(2010)Kesselheim, Sega, and Holm]{Kesselheim2010}
S.~Kesselheim, M.~Sega, and C.~Holm.
\newblock The icc* algorithm: A fast way to include dielectric boundary effects
  into molecular dynamics simulations.
\newblock \emph{arXiv:1003.1271}, 2010.

\bibitem[Arnold et~al.(2013)Arnold, Lenz, Kesselheim, Weeber, Fahrenberger,
  Roehm, Košovan, and Holm]{Arnold2013}
Axel Arnold, Olaf Lenz, Stefan Kesselheim, Rudolf Weeber, Florian Fahrenberger,
  Dominic Roehm, Peter Košovan, and Christian Holm.
\newblock Espresso 3.1: Molecular dynamics software for coarse-grained models.
\newblock volume 89 LNCSE, 2013.
\newblock \doi{10.1007/978-3-642-32979-1_1}.

\bibitem[Reed et~al.(2007)Reed, Lanning, and Madden]{Reed2007}
Stewart~K. Reed, Oliver~J. Lanning, and Paul~A. Madden.
\newblock Electrochemical interface between an ionic liquid and a model
  metallic electrode.
\newblock \emph{Journal of Chemical Physics}, 126, 2007.
\newblock ISSN 00219606.
\newblock \doi{10.1063/1.2464084}.

\bibitem[Rostami et~al.(2016)Rostami, Ghasemi, and
  Nedaaee~Oskoee]{rostami2016highly}
Samare Rostami, S~Alireza Ghasemi, and Ehsan Nedaaee~Oskoee.
\newblock A highly accurate and efficient algorithm for electrostatic
  interactions of charged particles confined by parallel metallic plates.
\newblock \emph{The Journal of chemical physics}, 145\penalty0 (12):\penalty0
  124118, 2016.

\bibitem[Biagooi et~al.(2020)Biagooi, Samanipour, Ghasemi, and
  Oskoee]{Biagooi2020}
Morad Biagooi, Mohammad Samanipour, S.~Alireza Ghasemi, and Seyedehsan~Nedaaee
  Oskoee.
\newblock Caviar: A simulation package for charged particles in environments
  surrounded by conductive boundaries.
\newblock \emph{AIP Advances}, 10, 2020.
\newblock ISSN 21583226.
\newblock \doi{10.1063/1.5140052}.

\bibitem[Shan et~al.(2020)Shan, Tang, Dang, Li, Yang, Xu, and Wu]{shan2020}
Tao Shan, Wei Tang, Xunwang Dang, Maokun Li, Fan Yang, Shenheng Xu, and Ji~Wu.
\newblock Study on a fast solver for poisson’s equation based on deep
  learning technique.
\newblock \emph{IEEE Transactions on Antennas and Propagation}, 68\penalty0
  (9):\penalty0 6725--6733, 2020.
\newblock \doi{10.1109/TAP.2020.2985172}.

\bibitem[Raissi et~al.(2019)Raissi, Perdikaris, and Karniadakis]{Raissi2019}
M.~Raissi, P.~Perdikaris, and G.~E. Karniadakis.
\newblock Physics-informed neural networks: A deep learning framework for
  solving forward and inverse problems involving nonlinear partial differential
  equations.
\newblock \emph{Journal of Computational Physics}, 378, 2019.
\newblock ISSN 10902716.
\newblock \doi{10.1016/j.jcp.2018.10.045}.

\bibitem[Breiman(2001)]{Breiman2001}
Leo Breiman.
\newblock Random forests.
\newblock \emph{Machine Learning}, 45, 2001.
\newblock ISSN 08856125.
\newblock \doi{10.1023/A:1010933404324}.

\bibitem[Pedregosa et~al.(2011)Pedregosa, Varoquaux, Gramfort, Michel, Thirion,
  Grisel, Blondel, Prettenhofer, Weiss, Dubourg, Vanderplas, Passos,
  Cournapeau, Brucher, Perrot, and Édouard Duchesnay]{Pedregosa2011}
Fabian Pedregosa, Gael Varoquaux, Alexandre Gramfort, Vincent Michel, Bertrand
  Thirion, Olivier Grisel, Mathieu Blondel, Peter Prettenhofer, Ron Weiss,
  Vincent Dubourg, Jake Vanderplas, Alexandre Passos, David Cournapeau,
  Matthieu Brucher, Matthieu Perrot, and Édouard Duchesnay.
\newblock Scikit-learn: Machine learning in python.
\newblock \emph{Journal of Machine Learning Research}, 12, 2011.
\newblock ISSN 15324435.

\bibitem[Abadi(2016)]{Abadi2016}
Martín Abadi.
\newblock Tensorflow: learning functions at scale.
\newblock \emph{ACM SIGPLAN Notices}, 51, 2016.
\newblock ISSN 0362-1340.
\newblock \doi{10.1145/3022670.2976746}.

\bibitem[Chollet(2015)]{Chollet2015}
François Chollet.
\newblock Keras: The python deep learning library.
\newblock \emph{Keras.Io}, 2015.

\bibitem[Walt et~al.(2011)Walt, Colbert, and Varoquaux]{Van}
Stéfan Van~Der Walt, S.~Chris Colbert, and Gaël Varoquaux.
\newblock The numpy array: A structure for efficient numerical computation.
\newblock \emph{Computing in Science and Engineering}, 13, 2011.
\newblock ISSN 15219615.
\newblock \doi{10.1109/MCSE.2011.37}.

\bibitem[Goodfellow et~al.(2016)Goodfellow, Bengio, and
  Courville]{Goodfellow-et-al-2016}
Ian Goodfellow, Yoshua Bengio, and Aaron Courville.
\newblock \emph{Deep Learning}.
\newblock MIT Press, 2016.
\newblock \url{http://www.deeplearningbook.org}.

\bibitem[a.~K.~Connect et~al.(1992)a.~K.~Connect, a.~Krogh, and
  a.~Hertz]{Connect1992}
a.~K.~Connect, a.~Krogh, and J.~a.~Hertz.
\newblock A simple weight decay can improve generalization.
\newblock \emph{Advances in Neural Information Processing Systems}, 4, 1992.

\bibitem[Kag et~al.(2022)Kag, Seshasayanan, and Gopinath]{kag2022physics}
Vijay Kag, Kannabiran Seshasayanan, and Venkatesh Gopinath.
\newblock Physics and data informed neural networks for two-dimensional
  turbulence.
\newblock \emph{arXiv preprint arXiv:2203.02555}, 2022.

\bibitem[Liu and Nocedal(1989)]{liu1989limited}
Dong~C Liu and Jorge Nocedal.
\newblock On the limited memory bfgs method for large scale optimization.
\newblock \emph{Mathematical programming}, 45\penalty0 (1):\penalty0 503--528,
  1989.

\end{thebibliography}






\end{document}